\documentclass[12pt]{iopart}
\usepackage{iopams}
\usepackage{graphicx}
\expandafter\let\csname equation*\endcsname\relax
\expandafter\let\csname endequation*\endcsname\relax
\usepackage{amsmath}
\usepackage{amssymb}
\usepackage{cite}

\begin{document}

\title[First passage times of random walks on random regular graphs]
{Analytical results for the distribution \\ 
of first-passage times of random walks \\ 
on random regular graphs
}

\author{Ido Tishby, Ofer Biham and Eytan Katzav}
\address{Racah Institute of Physics, 
The Hebrew University, Jerusalem 9190401, Israel.}
\eads{\mailto{ido.tishby@mail.huji.ac.il}, \mailto{ofer.biham@mail.huji.ac.il}, 
\mailto{eytan.katzav@mail.huji.ac.il}}

\begin{abstract}

We present analytical results for the distribution of
first-passage (FP) times of
random walks (RWs) on random regular graphs 
that consist of $N$ nodes of degree $c \ge 3$.
Starting from a random initial node at time $t=0$, 
at each time step $t \ge 1$ an RW hops 
into a random neighbor of its previous node.
In some of the time steps the RW may 
hop into a yet-unvisited node while in other time steps it may
revisit a node that has already been visited before.
We calculate
the distribution 
$P( T_{\rm FP} = t )$
of first-passage times
from a random initial node $i$ to a random target node $j$, where $j \ne i$.
We distinguish between FP trajectories whose backbone follows the shortest path
(SPATH)
from the initial node $i$ to the target node $j$
and FP trajectories whose backbone does not follow the shortest path
($\lnot {\rm SPATH}$).
More precisely, the SPATH trajectories 
from the initial node $i$ to the target node $j$
are defined as trajectories in which
the subnetwork that consists of the nodes
and edges along the trajectory is a tree network.
Moreover, the shortest path
between $i$ and $j$ on this subnetwork is the same as in the whole network.
The SPATH scenario is probable mainly 
when the length $\ell_{ij}$ of the shortest path between the initial node $i$ and the 
target node $j$ is small.
The analytical results are
found to be in very good agreement with the results 
obtained from computer simulations.

\end{abstract}


\noindent{\it Keywords}: 
Random network, 
random regular graph,
random walk, 
first-passage time,
shortest path,
backtracking,
retroceding.
 
\maketitle

\section{Introduction}

Random walk (RW) models 
were studied extensively 
in different geometries,
including
continuous space
\cite{Lawler2010b}, 
regular lattices 
\cite{Lawler2010a},
fractals 
\cite{ben-Avraham2000}
and 
random networks
\cite{Noh2004,Masuda2017}.
In the context of random networks
\cite{Havlin2010,Newman2010}, 
random walks 
provide useful tools for the analysis of
dynamical processes such as the
spreading of rumours, opinions and infections
\cite{Pastor-Satorras2001,Barrat2012}.
Consider an RW on a random network.
Starting at time $t=0$ from a random initial node $i=x_0$, 
at each time step $t \ge 1$ it hops
randomly to one of the neighbors of its previous node.
The RW thus generates a trajectory of the form
$x_0 \rightarrow x_1 \rightarrow \dots \rightarrow x_t \rightarrow \dots$,
where $x_{t}$ is the node visited at time $t$.
In some of the time steps the RW hops into nodes that
have not been visited before, while
in other time steps it hops into nodes that have
already been visited at an earlier time.
The mean number $\langle S \rangle_t$ of distinct nodes visited by an RW 
on a random network 
in the first $t$ time steps 
was recently studied 
\cite{Debacco2015}.
It was found that 
in the infinite network limit it scales linearly with $t$, namely
$\langle S \rangle_t \simeq r t$, 
where the coefficient
$r<1$ depends on the network topology.
These scaling properties resemble those obtained
for RWs on high dimensional lattices as well as on Bethe lattices and Cayley trees
\cite{Masuda2004}.

For an RW starting from an initial node $i$, the
first-passage (FP) time $T_{\rm FP}$ 
from $i$ to a target node $j$ 
(where $j \ne i$)
is the first time at which the RW 
visits the target node $j$
\cite{Redner2001,Finch2003}.
The first-passage time varies between different instances of the RW
trajectory and its properties can be captured by a suitable distribution.
The distribution of first-passage times may depend on the choice of the initial
node $i$ and the target node $j$.
In particular, it may depend on the 
length $\ell_{ij}$ of the shortest path
(also referred to as the distance) between $i$ and $j$.
Averaging over all possible choices of the initial node $i$ and the target node $j$
one obtains the distribution of first-passage times $P(T_{\rm FP}=t)$.
This problem was studied extensively on regular lattices
\cite{Redner2001,Finch2003}.
The mean first-passage time $\langle T_{\rm FP} \rangle$
of an RW on a random network was studied in Refs. 
\cite{Sood2005,Baronchelli2006,Lau2010,Bartolucci2021}.
The statistical properties of first passage processes in empirical networks
were applied in order to characterize the heterogeneity and correlations in
these networks 
\cite{Bassolas2021}.
Such analyses rely on the whole distribution of first passage times
in the network and not only on the mean first passage time.
However, closed-form analytical expressions for the distribution
$P(T_{\rm FP}=t)$ of first-passage times in random networks
have not been derived.

The special case 
in which the initial node $i$ is also chosen as the target node
is called the first return (FR) problem.
The distribution $P(T_{\rm FR}=t)$ of first return times was studied on the
Bethe lattice, which exhibits a tree structure of an infinite size
\cite{Hughes1982,Cassi1989,Giacometti1995}.
In a recent paper we presented analytical results for 
the distribution of  
first return times 
of RWs on random regular graphs (RRGs) consisting of $N$ nodes of degree $c \ge 3$
\cite{Tishby2021b}.
We considered
separately the scenario in which the RW returns
to the initial node $i$ by retroceding (RETRO) its own steps and
the scenario in which it does not retrocede its steps 
($\lnot {\rm RETRO}$) 
on the way back to $i$.
In the retroceding scenario an RW starting from the initial node $i$
eventually returns
to $i$ by stepping backwards via the same edges that it crossed
in the forward direction. This implies that in the retroceding scenario each edge
that belongs to the RW trajectory is crossed the same number of
times in the forward and backward directions.
In the non-retroceding scenario an RW starting from $i$ 
eventually returns to $i$ without retroceding its own steps.
This means that in 
the non-retroceding scenario the RW trajectory must include at least one cycle.
Using combinatorial and probabilistic methods we calculated the
conditional distributions of first return times,
$P(T_{\rm FR}=t | {\rm RETRO})$
and
$P(T_{\rm FR}=t | \lnot {\rm RETRO})$,
in the retroceding and non-retroceding scenarios, respectively.
We combined the results of the two scenarios with suitable weights and obtained
the overall distribution of first return times
$P(T_{\rm FR} = t)$
of RWs on RRGs of a finite size.

In this paper we present analytical results for 
the distribution of  
first-passage times 
of RWs on RRGs that consist of $N$ nodes of degree $c \ge 3$.
We consider separately the
case in which the first-passage trajectory from the initial node $i$ to the target node $j$ ($j \ne i$)
follows the shortest path (SPATH) between $i$ and $j$ and
the case in which it does not follow the shortest path ($\lnot {\rm SPATH}$).
In the SPATH trajectories the subnetwork that consists of the nodes
and edges along the trajectory is a tree network and the distance $\ell_{ij}$
between $i$ and $j$ on this subnetwork is the same as in the whole network.
In finite networks the SPATH scenario takes place mainly for pairs of nodes
for which the distance $\ell_{ij}$ is small.
We combine the results of the two cases with suitable weights and obtain the overall distribution
of first-passage times
$P(T_{\rm FP} = t)$
of RWs on RRGs.
The analytical results are
found to be in very good agreement with the results 
obtained from computer simulations.

The paper is organized as follows.
In Sec. 2 we briefly describe the random regular graph.
In Sec. 3 we present the random walk model.
In Sec. 4 we calculate the distribution of first-passage times
via SPATH trajectories.
In Sec. 5 we calculate the distribution of first-passage times via
non-SPATH trajectories.
The overall distribution of first passage times is calculated in Sec. 6.
The results are summarized and discussed in Sec. 7.
In Appendix A we calculate the number of distinct RW trajectories
between a pair of random nodes that are at a distance $\ell$ apart.
In Appendix B we evaluate useful sums that 
are used in the calculation of the moments of the
distribution of first-passage times.

\section{The random regular graph}

A random network (or graph) consists of a set of $N$ nodes that
are connected by edges in a way that is determined by some
random process.
For example, in a configuration model network the degree of each node is 
drawn independently from a given degree distribution $P(k)$ and
the connections are random and uncorrelated
\cite{Molloy1995,Molloy1998,Newman2001}.
The RRG is a special case of a configuration 
model network, in which the degree distribution is a degenerate
distribution of the form 
$P(k)=\delta_{k,c}$, namely
all the nodes are of the same degree $c$.
Here we focus on the case of $3 \le c \le N-1$,
in which for a sufficiently large value of $N$ the RRG consists of a single connected component
\cite{Bollobas2001}.
RRGs exhibit a local tree-like structure,
while at larger scales there is a broad spectrum of cycle lengths
\cite{Bonneau2017}.
In that sense RRGs differ from Cayley trees, which maintain their
tree structure by reducing the most peripheral nodes to leaf nodes of degree $1$.

The distribution of shortest path lengths (DSPL) of RRGs 
was studied in Refs. 
\cite{Hofstad2005,Nitzan2016}.
It was shown that the DSPL
follows a discrete Gompertz
distribution 
\cite{Gompertz1825,Shklovskii2005},
whose tail distribution is given by

\begin{equation}
P(L> \ell) = \exp \left[ - \eta \left( e^{b \ell} - 1 \right) \right],
\label{eq:taildist}
\end{equation}

\noindent
where 

\begin{equation}
\eta=\left( \frac{c}{ c-2 } \right) \frac{1}{ N },
\label{eq:eta}
\end{equation}

\noindent
and 

\begin{equation}
b=\ln(c-1).
\label{eq:b}
\end{equation}

\noindent
The probability mass function of the DSPL is given by

\begin{equation}
P(L=\ell) = P(L > \ell - 1) - P(L > \ell).
\label{eq:PL}
\end{equation}

Consider a pair of nodes $i$ and $j$ that reside at a distance $\ell$ from each other. 
This means that $i$ and $j$ are connected by at least one path of length $\ell$
and that there is no path connecting them which is shorter than $\ell$.
For some pairs of nodes the shortest path connecting them is unique.
For other pairs of nodes the shortest path between them may be degenerate, namely they may
be connected by several paths of length $\ell$ (and no paths shorter than $\ell$).
The degenerate shortest paths may include overlapping segments.
Thus, two shortest paths are considered distinct if they differ by at least one node.
The distribution of first-passage times (in the SPATH scenario)
from an initial node $i$ to a target node $j$
depends both on the shortest path length (or distance) $\ell$ between them and
on its degeneracy.
The probability that the first-passage trajectory of an RW from the initial node $i$ to 
a target node $j$ ($j \ne i$) will follow a shortest
path is proportional to the expectation value ${\mathbb E}[G|L=\ell]$ of the
number of degenerate shortest paths between $i$ and $j$. 
The degeneracy results from branching of the shortest path, which may 
occur at any point along the path.
The SPATH scenario is probable mainly when the distance between $i$ and $j$ is small.
Therefore, we can  
use an approximation in which we take into account only the degeneracy due to branching that occurs
in the first shell around the initial node $i$.
In this approximation, the degeneracy is given by the number of neighbors 
of $i$ that reside on
a shortest path from $i$ to $j$.
These neighbors of $i$ are at a distance $\ell-1$ from $j$,
where $\ell$ is the distance between $i$ and $j$. 
The degeneracy is equal to the number of possible first steps
of an RW starting from $i$, which reside on one of the shortest paths to $j$.
The degeneracy may take values in the range between $1$ and $c$.
The expectation value of the degeneracy of the shortest path
between the initial node $i$ and the target node $j$, 
which are at a distance $\ell \ge 2$ from each other,
can be approximated by

\begin{equation}
{\mathbb E}[G|L=\ell] \simeq 1 + (c-1)  \left[ 1 - \widetilde P(L>\ell-1|L>\ell-2) \right],
\label{eq:GL}
\end{equation}
 
\noindent
where

\begin{equation}
\widetilde P(L>\ell-1|L>\ell-2) =
\exp \left[ - \frac{(c-1)^{\ell-1}  }{N} \right]  
\label{eq:Ptilde}
\end{equation}

\noindent
is the probability that the distance between a node $i'$ selected via a random edge and
another random node $j$ is larger than $\ell-1$, given that it is larger than $\ell-2$
on the reduced network from which that edge is removed
\cite{Nitzan2016,Bonneau2017,Tishby2022}.

Below we explain the terms on the right hand side of Eq. (\ref{eq:GL}),
which approximates the expectation value ${\mathbb E}[G|L=\ell]$ 
of the number of degenerate shortest paths from $i$ to $j$ by
the number of neighbors of $i$ that reside on shortest paths to $j$, given that the distance
between $i$ and $j$ is equal to $\ell$. Under this condition, there is at least one
shortest path of length $\ell$ from $i$ to $j$, thus there is at least one neighbor
of $i$ which is at a distance $\ell-1$ from $j$. The first term on the right hand side
of Eq. (\ref{eq:GL}) accounts for this neighbor. 
However, for each one of the other $c-1$ neighbors of $i$ 
there is a non-zero probability that it may also reside at a distance
$\ell-1$ from $j$. Since the distance between $i$ and $j$ is $\ell$, the
length of the shortest path between any neighbor of $i$ and $j$ must be
larger than $\ell-2$. Therefore, for each one of the other $c-1$ neighbors of $i$,
the probability that it is at a distance $\ell-1$ from $j$ is given by
%
$\widetilde P(L=\ell-1|L>\ell-2) = 1 - \widetilde P(L>\ell-1|L>\ell-2)$.
%
Thus, the second term on the right hand side of Eq. (\ref{eq:GL}) accounts for the 
expected number of neighbors of $i$ that reside at a distance $\ell-1$ from $j$,
apart from the neighbor accounted for in the first term.
Overall, the right hand side of Eq. (\ref{eq:GL}) provides the expected number of 
neighbors of $i$ that reside on shortest paths from $i$ to $j$. 

In the special case of $\ell=1$ 
the expected number of degenerate shortest paths satisfies ${\mathbb E}[G|L=1] = 1$, 
due to the fact that in a simple
graph single edges cannot be degenerate.
The expectation value of the degeneracy increases as the path length $\ell$ is increased.
In case that $2  \le  \ell   \ll   \ln N/\ln (c-1)   $, 
Eq. (\ref{eq:GL}) can be approximated by

\begin{equation}
{\mathbb E}[G|L=\ell] \simeq 1 + \frac{ (c-1)^{\ell} }{N}.
\label{eq:GL2}
\end{equation}

\begin{figure}
\centerline{
\includegraphics[width=13.0cm]{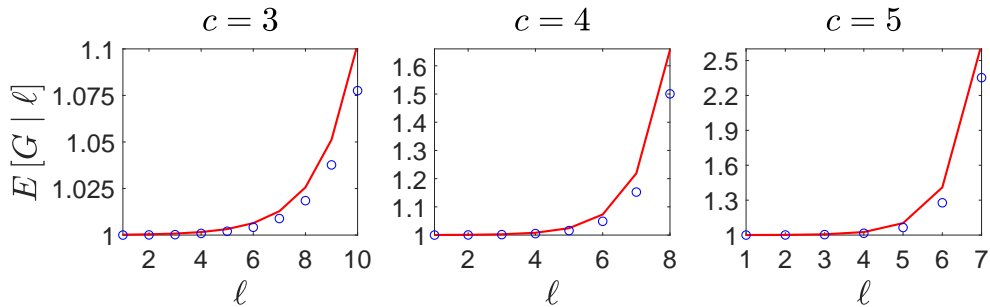} 
}
\caption{
Analytical results (solid lines) for the
expectation value of the degeneracy ${\mathbb E}[G|L=\ell]$
of shortest paths between pairs of nodes $i$ and $j$, as a function of the distance $\ell$ between them,
for an RRG that consists of $N=10^4$ nodes of degree $c=3$ (left), $c=4$ (center) and $c=5$ (right),
obtained from Eq. (\ref{eq:GL2}).
The results are found to be in good agreement with 
the results obtained from computer simulations (circles).
}
\label{fig:1}
\end{figure}

In Fig. \ref{fig:1} we present analytical results (solid lines) for the
expectation value of the degeneracy ${\mathbb E}[G|L=\ell]$
of shortest paths between pairs of nodes
$i$ and $j$ as a function of the distance $\ell$ between them
for $c=3$ (left), $c=4$ (center) and $c=5$ (right),
obtained from Eq. (\ref{eq:GL2}). 
The results are found to be in good agreement with 
the results obtained from computer simulations (circles).

A convenient way to construct an RRG
of size $N$ and degree $c$
($Nc$ must be an even number)
is to prepare the $N$ nodes such that each node is 
connected to $c$ half edges or stubs
\cite{Newman2010}.
At each step of the construction, one connects a pair of random stubs that 
belong to two different nodes $i$ and $j$ 
that are not already connected,
forming an edge between them.
This procedure is repeated until all the stubs are exhausted.
The process may get stuck before completion whenever
all the remaining stubs belong to the
same node or to pairs of nodes that are already connected.
In such case one needs to perform some random reconnections
in order to complete the construction.

\section{The random walk model}

\begin{figure}
\centerline{
\includegraphics[width=5.0cm]{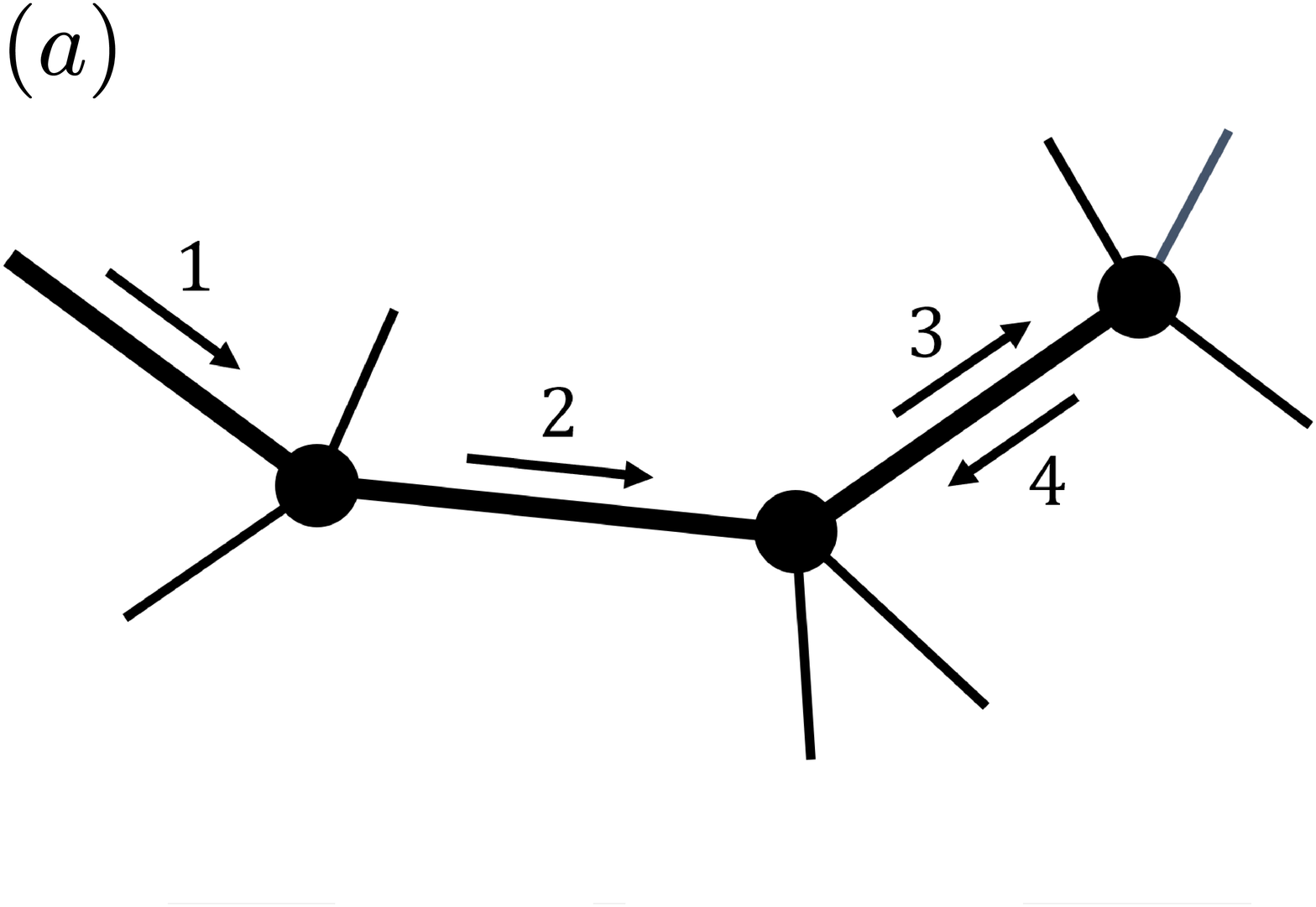}
}
\centerline{
\includegraphics[width=5.0cm]{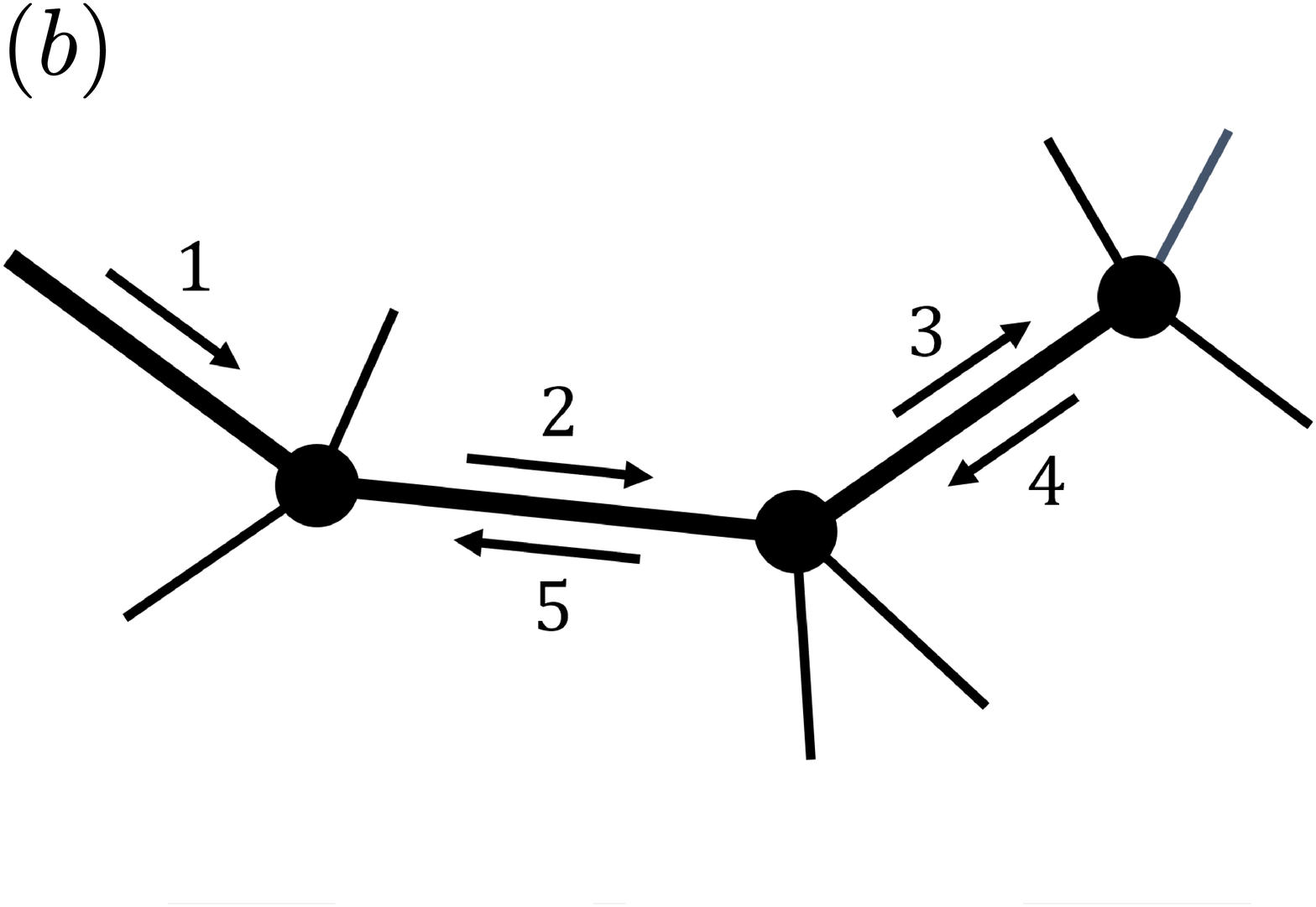}
}
\centerline{
\includegraphics[width=5.0cm]{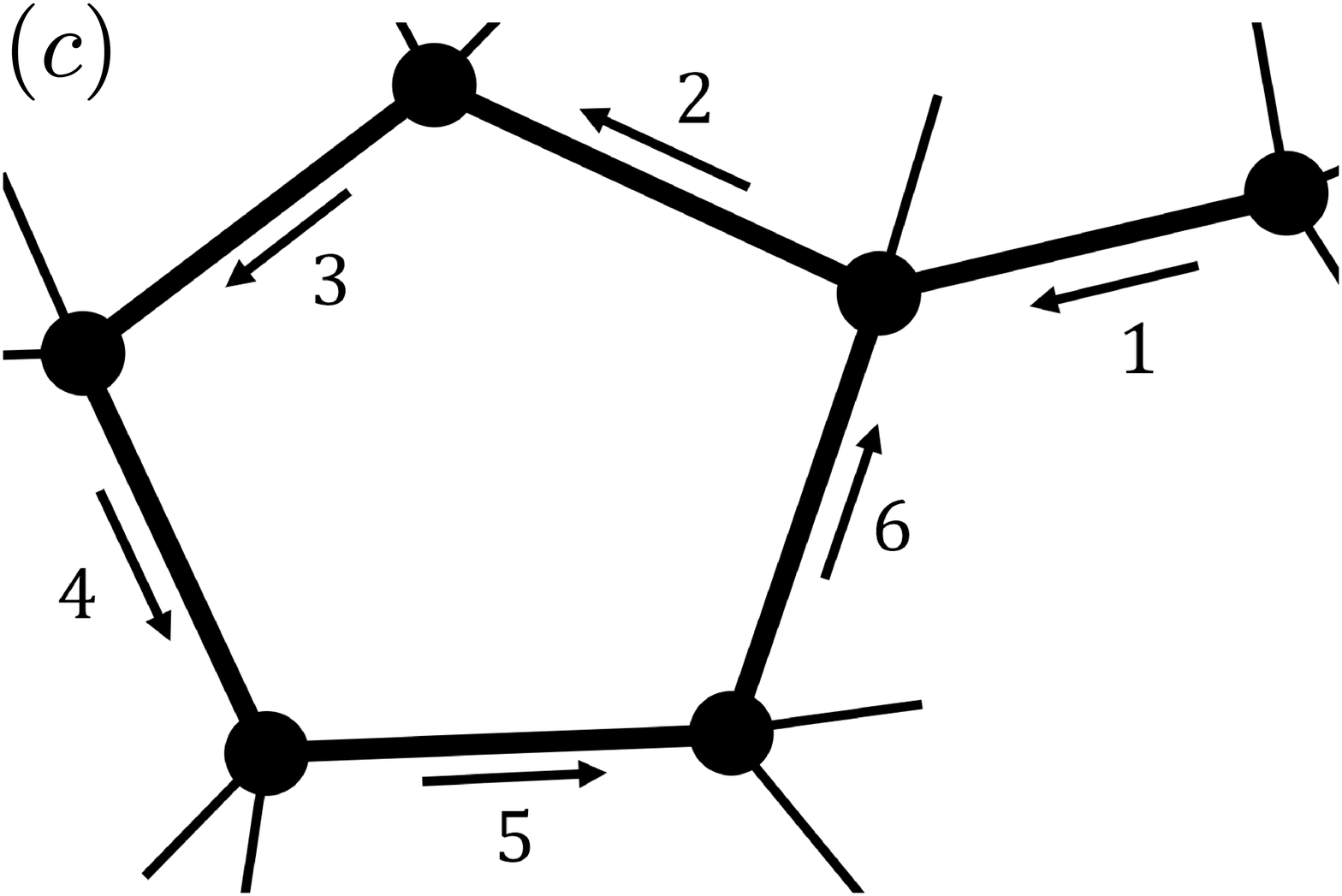}
}
\caption{
Schematic illustrations of possible events taking place along the path 
of an RW on an RRG:
(a) a path segment that includes a backtracking step
into the previous node (step no. 4);
(b) a path segment that includes a backtracking step (step no. 4), which is 
followed by a retroceding step (step no. 5);
(c) a path that includes a retracing step (step no. 6) in which the RW hops into a node that was
already visited a few time steps earlier. Retracing steps are not possible in the infinite
network limit. They take place only in finite RRGs, which include cycles. 
Note that in this illustration the RRG
is of degree $c=4$.
}
\label{fig:2}
\end{figure}

Consider an RW on an RRG that consists of $N$ nodes of degree $c \ge 3$. 
Starting from a random initial node at time $t=0$, 
at each time step $t \ge 1$ the RW hops   
into a random neighbor of its previous node.
The probability to hop into each one of these neighbors is $1/c$.
For sufficiently large $N$ the RRG consists of a single connected
component, thus an RW starting from any initial node
may eventually reach any other node in the network.
In some of the time steps an RW may visit nodes that have not been visited before
while in other time steps it may revisit nodes that have already been visited before. 
For example, at each time step $t \ge 3$ the RW may backtrack into the previous node with 
probability of $1/c$.
In the infinite network limit the RRG exhibits a tree structure.
Therefore, in this limit the backtracking mechanism is the only way in which an RW
may hop from a newly visited node to a node that has already been visited before.
Such backtracking step may be followed by retroceding steps in which the RW
continues to go backwards along its own path.
However, in finite networks the RW may also utilize cycles to retrace its path 
and revisit nodes it has already visited three or more time steps earlier.
In Fig. \ref{fig:2} we present a
schematic illustration of some of the events that may take place along the path 
of an RW on an RRG.
In Fig. \ref{fig:2}(a) we show a path segment that includes a backtracking step,
in which the RW moves back into the previous node (step no. 4). 
In Fig. \ref{fig:2}(b) we show a path segment that includes a backtracking step
(step no. 4)
which is followed by a retroceding step (step no. 5).
In Fig. \ref{fig:2}(c) we show a path segment that includes a retracing step (step no. 6),
in which the RW enters a node that was already visited five time steps earlier.

\section{The distribution of first-passage times via SPATH trajectories}

Consider an RW 
on an RRG of a finite size $N$,
starting at $t=0$ from an initial node $i$.
The time at which the RW visits a given target node $j$
($j \ne i$)
for the first time
is called the first-passage time from $i$ to $j$.
We first consider the case in which the initial node $i$ is random and the target node $j$ is
a random node that resides
at a distance $\ell$ from $i$, where $\ell \ge 1$.
Clearly, for $t < \ell$ the first-passage probability satisfies
$P(T_{\rm FP}=t|\ell)=0$.
For $t \ge \ell$ the first-passage probability satisfies
$P(T_{\rm FP}=t|\ell) > 0$
and for any finite network

\begin{equation}
\sum_{t=\ell}^{\infty} 
P(T_{\rm FP}=t|\ell) = 1.
\end{equation}

\noindent
The probability that an
RW trajectory will follow the shortest path from the initial node $i$ to the target node $j$ 
(or one of the shortest paths in case they are degenerate)
is denoted by $P({\rm SPATH}|\ell)$.
In the simplest trajectory of this type the RW moves forward along a shortest path from $i$ to $j$
at all time steps, such that the first-passage time is $t=\ell$.
However, in some of the time steps the RW may backtrack its path and possibly retrocede,
resulting in a longer trajectory that still follows the shortest path. 
Furthermore, the RW may also step away from the shortest path and then return
to the same point along the shortest path
by a combination of backtracking and retroceding steps.
Such trajectories are included in the SPATH scenario.
More formally, for a first-passage trajectory to be an SPATH trajectory it
must satisfy two conditions: (a)
the subnetwork that consists of the nodes
and edges along the trajectory is a tree network; (b) the distance  
between $i$ and $j$ on this subnetwork is the same as the distance $\ell_{ij}$ in the whole network.
First-passage trajectories that do not satisfy one or both of these conditions are
non-SPATH trajectories.

\begin{figure}
\centerline{
\includegraphics[width=5cm]{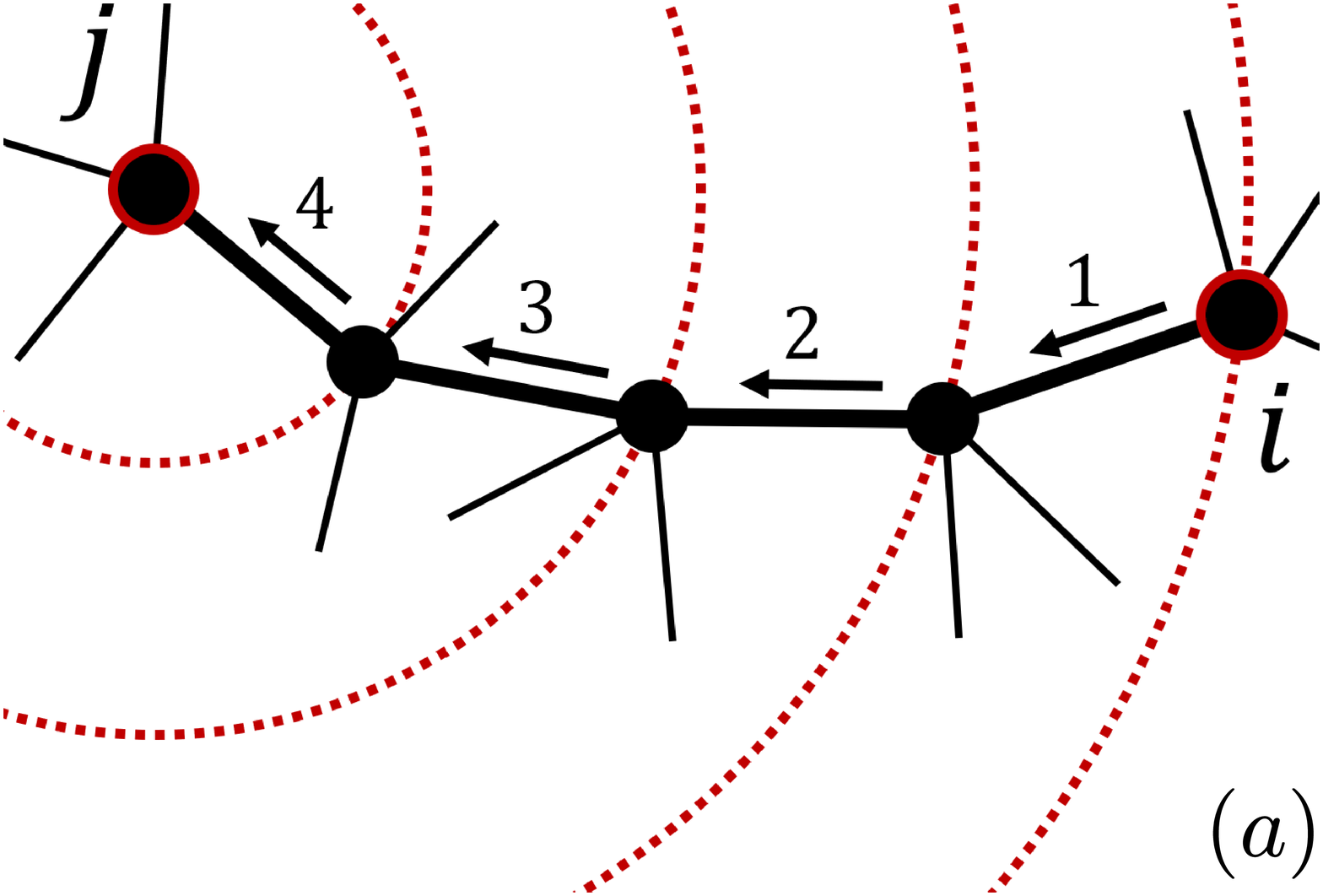}
}
\centerline{
\includegraphics[width=5cm]{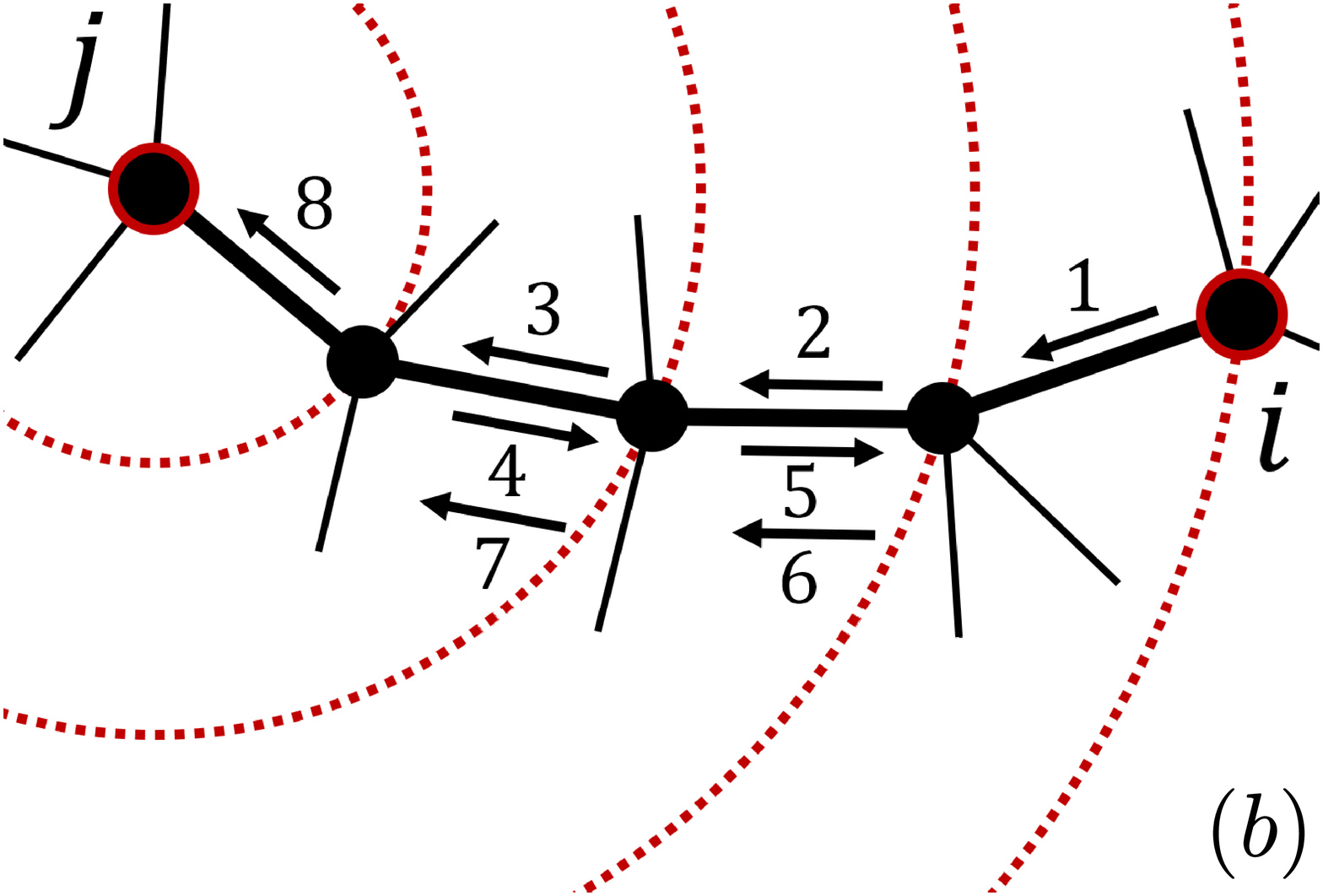}
}
\centerline{
\includegraphics[width=5cm]{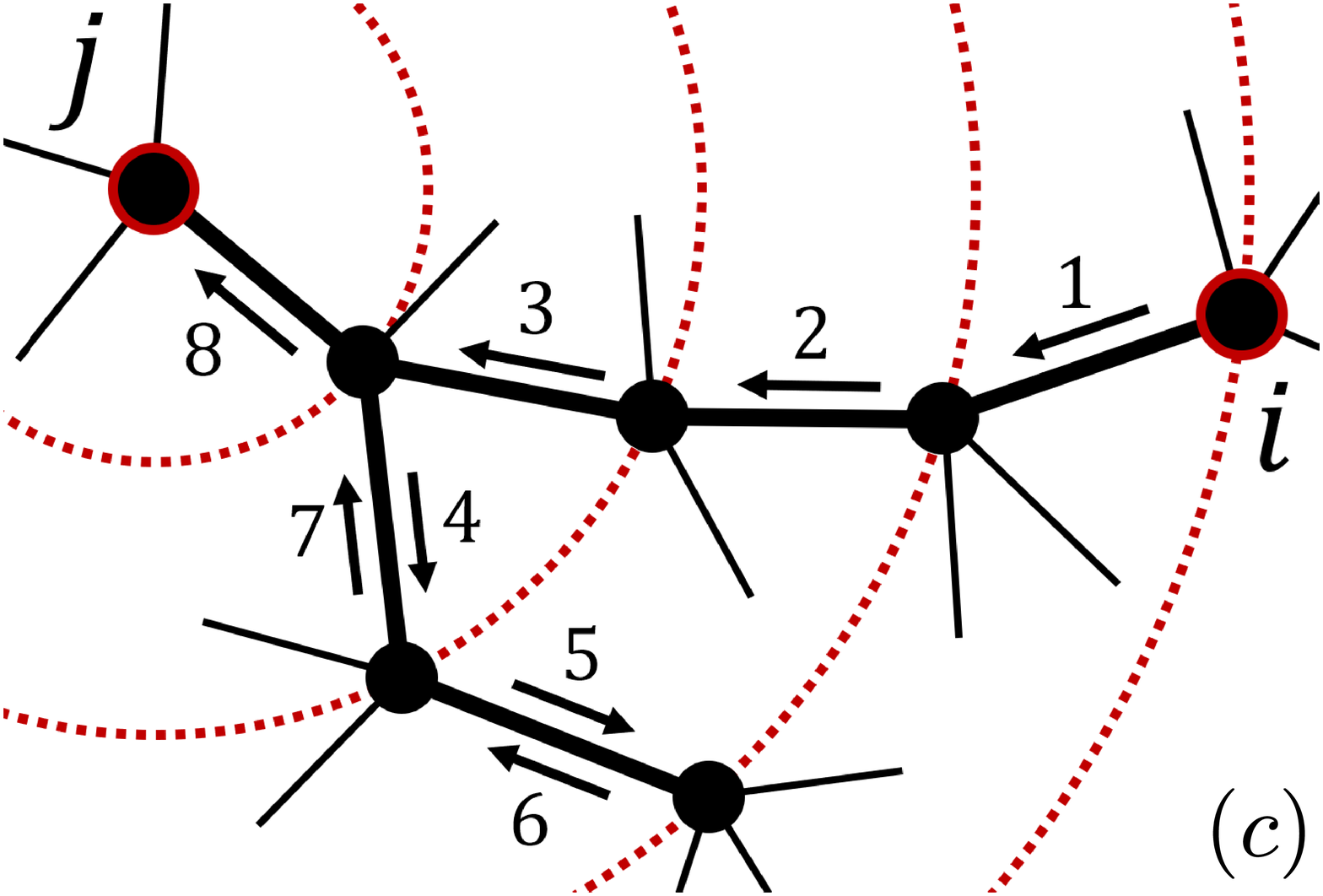}
}
\caption{
Schematic illustration of first passage RW trajectories 
from node $i$ to node $j$ that follow the shortest path 
between $i$ and $j$:
(a) an RW trajectory that goes along the shortest path without
any backtracking moves or diversions;
(b) an RW trajectory that includes backtracking and retroceding
steps along the shortest path;
(c) an RW trajectory that leaves the
shortest path and then returns by
retroceding its steps. It then follows the shortest
path towards $j$.
In this illustration the RRG is of degree $c=4$ 
and the shortest path length between $i$ and $j$ 
is $\ell_{ij}=4$.
} 
\label{fig:3}
\end{figure}

In Fig. \ref{fig:3} we present a
schematic illustration of first-passage RW trajectories 
from the initial node $i$ to the target node $j$ that follow the shortest path 
from $i$ to $j$.
The simplest among these paths is
an RW trajectory that goes along the shortest path without
any backtracking moves or diversions
[Fig. \ref{fig:3}(a)].
Such RW trajectories may also include backtracking and retroceding
steps along the shortest path [Fig. \ref{fig:3}(b)]
as well as diversion segments in which 
the RW leaves the
shortest path and then returns to the same node by
retroceding its steps and then continues to follow the shortest
path towards the target node $j$
[Fig. \ref{fig:3}(c)].

The probability that 
an RW trajectory starting from 
a random initial node
$i$ will follow a 
shortest path (SPATH) from $i$ to a random target node $j$ 
that resides at a distance $\ell$ from $i$,
and that the first-passage time will be $t$ is denoted by
$P(T_{\rm FP}=t, {\rm SPATH} | \ell)$.
Clearly, for $t < \ell$ it satisfies
$P(T_{\rm FP}=t, {\rm SPATH} | \ell)=0$.
For $t \ge \ell$, 
this probability
is given by

\begin{equation}
P(T_{\rm FP}=t, {\rm SPATH} | \ell) =
\left\{
\begin{array}{ll}
{\mathbb E}[G|\ell] B(t,\ell) \left( \frac{1}{c} \right)^t & \ \ \ \ \ \  t-\ell  \ \   {\rm even} \\
0 & \ \ \ \ \ \  t-\ell  \ \ \ \  {\rm odd},
\end{array}
\right.  
\label{eq:PTfpSell}
\end{equation}

\noindent
where 
${\mathbb E}[G|\ell]$ is the expected number of degenerate shortest paths 
of length $\ell$.
The coefficient $B(t,\ell)$ is a combinatorial factor 
that accounts for the number of distinct RW
trajectories that follow a given shortest path
from $i$ to a target node $j$, which resides at a distance $\ell$ from $i$,
and reach $j$ for the first time after $t$ time steps.
The term $(1/c)^t$ accounts for the probability that an RW will follow any
specific trajectory of $t$ time steps. This is due to the fact that at each time step
the probability that the RW will hop to each neighbor of the current node
is $1/c$. The right hand side of Eq. (\ref{eq:PTfpSell}) is thus a product of
the total number of SPATH trajectories of $t$ time steps by the probability that an RW will
follow a given trajectory.

\begin{figure}
\centerline{
\includegraphics[width=8cm]{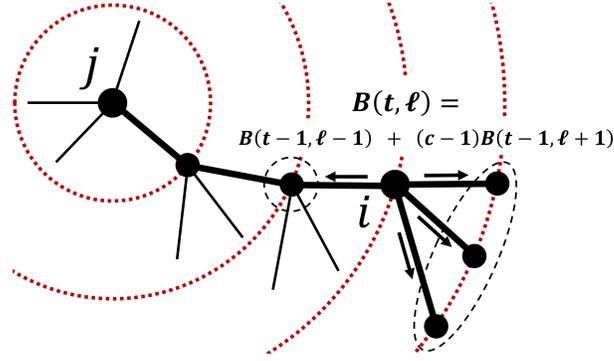}
}
\caption{
Illustration of the recursion equation (\ref{eq:Atell}) for the number
$B(t,\ell)$ of distinct RW trajectories that start from a random node $i$ and follow the
shortest path to some other node $j$, which resides at a distance $\ell$ from $i$.
In this illustration the RRG is of degree $c=4$.
}
\label{fig:4}
\end{figure}

We now focus on a single shortest path of length $\ell$ between a pair of nodes $i$ and $j$.
The coefficients $B(t,\ell)$ satisfy the recursion equations

\begin{equation}
B(t,\ell) = B(t-1,\ell-1) + (c-1) B(t-1,\ell+1),
\label{eq:Atell}
\end{equation}

\noindent
where the distance satisfies $\ell \ge 1$, the first-passage time satisfies $t \ge \ell$, and the
difference $t-\ell$ is an even number
[otherwise $B(t,\ell)=0$].
The recursion equation (\ref{eq:Atell}) is illustrated in Fig. \ref{fig:4},
depicting a pair of nodes $i$ and $j$, at a distance $\ell$ apart.
One neighbor of $i$ resides along the shortest path to $j$ and is 
thus at a distance $\ell-1$ from $j$, while the other $c-1$ neighbors
of $i$ are at a distance $\ell+1$ from $j$.
Therefore, an RW trajectory starting from $i$  
and ending up at $j$ may step at time $t=1$ either into the neighbor of $i$ that
resides along the shortest path to $j$ or into one of the other $c-1$ neighbors.
The boundary condition is $B(\ell,\ell)=1$ for $\ell \ge 1$.
In the special case of $\ell=1$ and $t \ge 3$, the recursion equation is
given by 

\begin{equation}
B(t,1)=(c-1) B(t-1,2).
\end{equation}

\noindent
Actually, the $B(t,\ell)$ coefficients are also defined for $\ell=0$.
In this case, they provide the number of RW trajectories that
return to the initial node for the first time at time $t$,
and are thus important in the analysis of the first return problem,
recently studied in Ref.
\cite{Tishby2021b}.

In Appendix A we present a solution of the recursion relations given by Eq. (\ref{eq:Atell})
for $\ell \ge 1$ and $t \ge \ell$
with the boundary conditions specified above.
It is found that for any distance $\ell \ge 1$
between the initial node $i$ and the target node $j$ 
and for any time $t \ge \ell$  
the number of distinct RW trajectories that follow a given shortest path 
from $i$ to $j$
and for which the first-passage time is $t$, is given by

\begin{equation}
B(t,\ell) =  
 \left\{
\begin{array}{ll}
\frac{\ell}{t} 
\binom{t}{   \frac{t+\ell}{2}   } 
(c-1)^{ \frac{t-\ell}{2} }  & \ \ \ \ \ \ \  t-\ell  \ \   {\rm even} \\
0 & \ \ \ \ \  \ \  t-\ell  \ \  {\rm odd},
\end{array}
\right.  
\label{eq:Btell}
\end{equation}

\noindent
where $\binom{n}{m}$ is the binomial coefficient.
The coefficient $B(t,\ell)$ captures the contribution of a single shortest path
to the distribution of first passage times via the SPATH scenario.

In order to proceed one needs to account for the contributions of all the
degenerate shortest paths between the initial node and the target node.
The probability that an RW starting from a random initial node $i$ 
will reach a random target node $j$ that resides at a distance $\ell$ from $i$  
for the first time via the SPATH scenario is given by

\begin{equation}
P({\rm SPATH} | \ell) = 
\sum_{t=\ell}^{\infty}
P(T_{\rm FP}=t, {\rm SPATH} | \ell).
\label{eq:PSell}
\end{equation}

\noindent
Inserting 
$P(T_{\rm FP} = t, {\rm SPATH} | \ell)$
from Eq. (\ref{eq:PTfpSell})
[with $B(t,\ell)$ from Eq. (\ref{eq:Btell})] 
into Eq. (\ref{eq:PSell}) and carrying out the summation,
we obtain

\begin{equation}
P({\rm SPATH} | \ell) = \left( \frac{1}{c-1} \right)^{\ell} {\mathbb E}[G|\ell].
\label{eq:PSell2}
\end{equation}

\noindent
The first term on the right hand side of Eq. (\ref{eq:PSell2}) 
accounts for the contribution of a single shortest path while
the second term accounts for the expected number of such paths.
Inserting 
the approximated result for the degeneracy
${\mathbb E}[G|\ell]$, given by Eq. (\ref{eq:GL2}),
into Eq. (\ref{eq:PSell2}),
we obtain

\begin{equation}
P({\rm SPATH} | \ell) \simeq \left( \frac{1}{c-1} \right)^{\ell}  + \frac{1}{N}.
\label{eq:PSell3}
\end{equation}

\noindent
The first term in Eq. (\ref{eq:PSell3}) accounts for the contribution of the 
single shortest path which is guaranteed to exist, given that the distance
between the initial node $i$ and the target node $j$ is equal to $\ell$.
The second term accounts for the expected contribution of degenerate shortest paths
of length $\ell$.

The complementary probability is given by

\begin{equation}
P(\lnot {\rm SPATH} | \ell) \simeq 1 - \left( \frac{1}{c-1} \right)^{\ell} - \frac{1}{N}.
\label{eq:PnoSell2}
\end{equation}

\begin{figure}
\centerline{
\includegraphics[width=13cm]{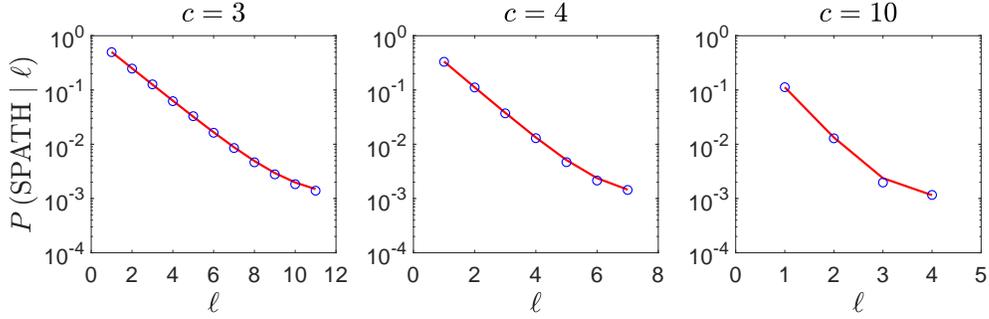}
}
\caption{
Analytical results (solid lines) for the probability $P({\rm SPATH}|\ell)$
that the first passage of an RW from a random initial node $i$ to a random
target node $j$, which is at a distance $\ell$ from $i$,
will occur via the ${\rm SPATH}$ scenario.
The results are presented for  
RRGs of size $N=1000$ and for $c=3$ (left), $c=4$ (center) and $c=10$ (right).
The analytical results, obtained from Eq. (\ref{eq:PSell3}) are in very good
agreement with the results obtained from computer simulations (circles).
}
\label{fig:5}
\end{figure}

In Fig. \ref{fig:5} we present
analytical results (solid lines) for the probability $P({\rm SPATH}|\ell)$
that the first-passage of an RW from a random initial node $i$ to a random
target node $j$ which is at a distance $\ell$ from $i$,
will take place via the ${\rm SPATH}$ scenario,
as a function of the distance $\ell$.
The results are shown for RRGs of size $N=1000$ and 
$c=3$ (left), $c=4$ (center) and $c=10$ (right).
The analytical results, obtained from Eq. (\ref{eq:PSell3}) are found to be in good
agreement with the results obtained from computer simulations (circles).

The distribution of first-passage times via SPATH trajectories,
from a random node $i$ to a random node $j$ that resides at a distance
$\ell$ from $i$, is given by

\begin{equation}
P(T_{\rm FP} = t| \ell, {\rm SPATH})
=
\frac{P(T_{\rm FP} =t, {\rm SPATH} | \ell)}{P({\rm SPATH} | \ell)}.
\label{eq:PTfp_ellS}
\end{equation}

\noindent
Plugging in $P(T_{\rm FP}=t, {\rm SPATH} | \ell)$ from Eq. (\ref{eq:PTfpSell})
and $P({\rm SPATH} | \ell)$ from Eq. (\ref{eq:PSell2})
into Eq. (\ref{eq:PTfp_ellS}), 
it is found that 
$P(T_{\rm FP} = t| \ell, {\rm SPATH})=0$
for $t < \ell$,
while for $t \ge \ell$ it is given by

\begin{equation}
P(T_{\rm FP} = t| \ell, {\rm SPATH})
=
\left\{
\begin{array}{ll}
 \frac{\ell}{t}
\binom{t}{ \frac{t+\ell}{2} }
\left( 1 - \frac{1}{c} \right)^{\frac{t+\ell}{2}}
\left( \frac{1}{c} \right)^{\frac{t-\ell}{2}} 
& \ \ \ \  t-\ell  \ \   {\rm even} \\
0 & \ \ \ \  t-\ell  \ \  {\rm odd}.  
\end{array}
\right.  
\label{eq:PTfp_ellS2}
\end{equation}

\noindent
The corresponding tail distribution is given by

\begin{equation}
P(T_{\rm FP} > t| \ell, {\rm SPATH}) =
\sum_{t'=t+1}^{\infty}
P(T_{\rm FP} = t'| \ell, {\rm SPATH}).
\label{eq:PTFPtellSP}
\end{equation}

\begin{figure}
\centerline{
\includegraphics[width=12cm]{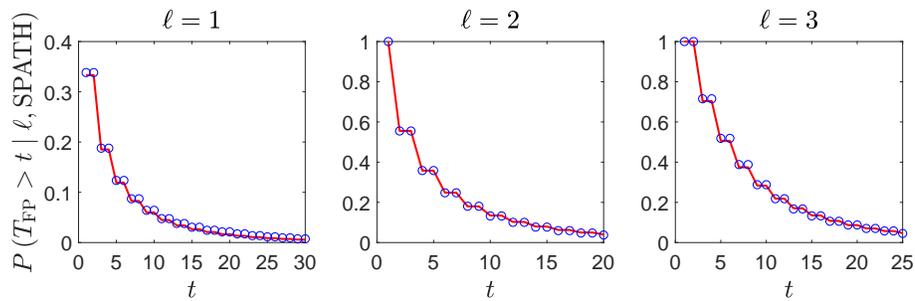}
}
\caption{
Analytical results (solid line) for the tail distribution
$P(T_{\rm FP}>t | \ell, {\rm SPATH})$
of first-passage times of RWs that follow the shortest
paths between pairs of nodes which are at a distance $\ell$ from
each other. The results are presented for RRGs of size $N=1000$ and degree $c=3$
and $\ell=1$ (left), $\ell=2$ (center) and $\ell=3$ (right).
The analytical results, obtained from Eq. (\ref{eq:PTFPtellSP}), are in very good
agreement with the results obtained from computer simulations (circles).
}
\label{fig:6}
\end{figure}

In Fig. \ref{fig:6} we present
analytical results (solid line) for the tail distribution
$P(T_{\rm FP}>t | \ell, {\rm SPATH})$
of first-passage times for RW trajectories that follow the shortest
path between pairs of nodes which are at a distance $\ell$ from
each other, for $\ell=1$ (left), $\ell=2$ (center) and $\ell=3$ (right).
The results are shown for RRGs of size $N=1000$ and degree $c=3$.
The analytical results, obtained from 
Eq. (\ref{eq:PTFPtellSP}) are in very good
agreement with the results obtained from computer simulations
(circles).

The generating function of the distribution
$P(T_{\rm FP} = t| \ell, {\rm SPATH})$
is given by

\begin{equation}
V_{\ell}(x) = \sum_{t=1}^{\infty}
x^t P(T_{\rm FP} = t| \ell, {\rm SPATH}).
\label{eq:GFfpt}
\end{equation}

\noindent
Such generating functions were studied extensively in the context of RWs on 
regular lattices
\cite{Finch2003}.
Inserting $P(T_{\rm FP} = t| \ell, {\rm SPATH})$
from Eq. (\ref{eq:PTfp_ellS2}) into Eq. (\ref{eq:GFfpt})
and carrying out the summation, we obtain

\begin{equation}
V_{\ell}(x) = 
\left( 1 - \frac{1}{c} \right)^{\ell} x^{\ell}
\, _2F_1 \left[ \left.
\begin{array}{c}
\frac{\ell}{2},  \frac{\ell+1}{2} \\
\ell+1
\end{array}
\right|    \frac{4(c-1)}{c^2} x^2
\right],
\label{eq:GFfpt2}
\end{equation}

\noindent
where  
$_2F_1 \left[ \left.
\begin{array}{c}
a, b \\
c
\end{array}
\right| z 
\right]$
is the hypergeometric function
\cite{Olver2010}.
Applying identity (\ref{eq:I}) 
from Appendix A
on the hypergeometric function
on the right hand side of Eq. (\ref{eq:GFfpt2}),
it can be simplified to 

\begin{equation}
V_{\ell}(x) = 
\left[ \frac{ c - \sqrt{c^2 - 4(c-1)x^2} }{2x} \right]^{\ell}. 
\label{eq:GFfpt3}
\end{equation}

The $r$'th 
moment of 
$P(T_{\rm FP} = t| \ell, {\rm SPATH})$
is given by

\begin{equation}
\mathbb{E}[T_{\rm FP}^r | \ell, {\rm SPATH}]
=
\sum_{t=1}^{\infty}
t^r P(T_{\rm FP}=t | \ell, {\rm SPATH}).
\label{eq:ETfpr}
\end{equation}

\noindent
Inserting $r=1$ in Eq. (\ref{eq:ETfpr}) we obtain the mean first-passage time
$\mathbb{E}[T_{\rm FP} | \ell, {\rm SPATH}]$.
It can also be obtained directly from the generating function
$V_{\ell}(x)$ by

\begin{equation}
\mathbb{E}[T_{\rm FP} | \ell, {\rm SPATH}]
=
\frac{d  V_{\ell}(x)}{dx } \bigg\vert_{x=1},
\label{eq:ETfprg2}
\end{equation}

\noindent
which yields

\begin{equation}
\mathbb{E}[T_{\rm FP} | \ell, {\rm SPATH}]
=
\frac{c}{c-2} \ell.
\label{eq:ETfp1p}
\end{equation}

\noindent
The second factorial moment is given by

\begin{equation}
\mathbb{E}[T_{\rm FP} (T_{\rm FP}-1) | \ell, {\rm SPATH}]
=
\frac{d^2  V_{\ell}(x)}{dx^2 } \bigg\vert_{x=1},
\label{eq:ETfprg2}
\end{equation}

\noindent
which yields

\begin{equation}
\mathbb{E}[T_{\rm FP} (T_{\rm FP}-1) | \ell, {\rm SPATH}]
=
\frac{c^2 \ell^2}{(c-2)^2}
- \frac{c^2-8c+8}{(c-2)^2} c \ell.
\label{eq:ETfprg3}
\end{equation}

\noindent
Combining the results for the first moment and the second factorial moment,
we obtain

\begin{equation}
\mathbb{E}[T_{\rm FP}^2 | \ell, {\rm SPATH}]
=
\frac{ 4 c (c-1) }{(c-2)^3} \ell
+
\left( \frac{c}{c-2} \right)^2 \ell^2.
\label{eq:ETfp2p}
\end{equation}

\noindent
Using the results presented above for the first and second moments we obtain the variance,
which is given by

\begin{equation}
{\rm Var}[T_{\rm FP} | \ell, {\rm SPATH}]
=
\frac{4 c (c-1)}{(c-2)^3} \ell.
\label{eq:VTFP1}
\end{equation}

The probability that the first-passage trajectory 
of an RW
from a random initial node $i$ to a random target
node $j$ in an RRG of a finite size $N$
will follow the shortest path from $i$ to $j$ is given by

\begin{equation}
P( {\rm SPATH} ) = 
\sum_{\ell=1}^{N-1}
P( {\rm SPATH}|\ell) P(L=\ell),
\label{eq:SP}
\end{equation}

\noindent
where 
$P( {\rm SPATH}|\ell)$ is given by Eq. (\ref{eq:PSell2}) and
$P(L=\ell)$ is given by Eqs. (\ref{eq:taildist})-(\ref{eq:PL}).
Using Eq. (\ref{eq:PL}), 
we obtain

\begin{equation}
P( {\rm SPATH} ) = 
\sum_{\ell=1}^{N-1}
P( {\rm SPATH}|\ell) \left[ P(L>\ell-1) - P(L>\ell) \right].
\label{eq:SP2}
\end{equation}

\noindent
Rearranging the summation indices, we obtain

\begin{equation}
P( {\rm SPATH} ) = 
P({\rm SPATH}|1) +
\sum_{\ell=1}^{N-2}
\left[ P({\rm SPATH}|\ell+1) - P({\rm SPATH}|\ell) \right]     P(L>\ell).
\label{eq:SP2}
\end{equation}

\noindent
The transformation from Eq. (\ref{eq:SP}) to Eq. (\ref{eq:SP2})
essentially amounts to a summation by parts.
Inserting $P({\rm SPATH}|\ell)$ from Eq. (\ref{eq:PSell2}) into Eq. (\ref{eq:SP2})
and rearranging terms, 
we obtain

\begin{equation}
P( {\rm SPATH} ) \simeq 
1   + \frac{1}{N}   - \frac{ c-2 }{c-1} 
\sum_{\ell=0}^{N-1}
\left( \frac{1}{c-1} \right)^{\ell} 
P(L>\ell),
\label{eq:PStail}
\end{equation}

\noindent
where we use the fact that $P(L>0)=1$.
Clearly, the upper limit of the summation in Eq. (\ref{eq:PStail}) can be replaced by $\infty$
with negligible effect on the result.
Using the definition of $b$ from
Eq. (\ref{eq:b}), we rewrite Eq. (\ref{eq:PStail}) in the form

\begin{equation}
P( {\rm SPATH} ) \simeq 
1 + \frac{1}{N} - \frac{ c-2 }{c-1} 
\sum_{\ell=0}^{\infty}
e^{- b \ell}
P(L>\ell).
\label{eq:PStail2}
\end{equation}

\noindent
It can also be written in the form

\begin{equation}
P(  {\rm SPATH} ) \simeq 
1 + \frac{1}{N} - \frac{ c-2 }{c-1} 
M_0
\label{eq:PStail3}
\end{equation}

\noindent
where

\begin{equation}
M_0 =
\mathcal{L}[P(L>\ell)](s=b) =
\sum_{\ell=0}^{\infty}
e^{- b \ell}
P(L>\ell) 
\label{eq:PStail4}
\end{equation}

\noindent
and $\mathcal{L}[P(L>\ell)](s)$
is the discrete Laplace transform 
(or the unilateral Z-transform)
of the tail distribution $P(L>\ell)$,
evaluated at $s=b$.
An explicit expression for $M_0$ is given by Eq. (\ref{eq:M_0a}).
It is based on a recent calculation of the discrete Laplace transform
$\mathcal{L}[P(L>\ell)](s)$
\cite{Tishby2022}.
In the large network limit it can be reduced to Eq. (\ref{eq:M_0b}).

Eqs. (\ref{eq:PStail})-(\ref{eq:PStail4}) provide an interesting relation between
the geometric properties of the network, captured by $P(L>\ell)$, and
the probability $P( {\rm SPATH} )$ which is a dynamical property of the RW
trajectories. It exemplifies the potential use of the DSPL in the analysis of
dynamical processes on networks.
Inserting $M_0$ from Eq. (\ref{eq:M_0b}) into Eq. (\ref{eq:PStail3})
and rearranging terms, we obtain

\begin{equation}
P({\rm SPATH}) \simeq \left( \frac{c}{c-1} \right) \frac{1}{N}
\left[ \frac{ \ln \left( \frac{ c-2 }{c} N \right) - \gamma + 1 }{\ln(c-1)}
+ \frac{ c^2 - 6c + 4 }{2c(c-2)} \right]
+ {\mathcal O} \left( \frac{1}{N^2} \right),
\label{eq:PStail5}
\end{equation}

\noindent
where $\gamma$ is the Euler-Mascheroni constant
\cite{Olver2010}.
Using the result for the mean distance $\langle L \rangle$,
presented in Ref. \cite{Tishby2022}, Eq. (\ref{eq:PStail5}) 
can be written in the form

\begin{equation}
P({\rm SPATH}) \simeq
\left( \frac{c}{c-1} \right) \frac{1}{N}
\left[ \langle L \rangle + \frac{1}{\ln(c-1)} - 2 \frac{c-1}{c(c-2)} \right]
+ \mathcal{O} \left( \frac{1}{N^2} \right).
\label{eq:PStail6}
\end{equation}

\noindent
In the limit of large and dense networks, Eq. (\ref{eq:PStail6}) can be simplified
to 

\begin{equation}
P({\rm SPATH}) \simeq
\left( \frac{c}{c-1} \right) \frac{\langle L \rangle}{N}
+ \mathcal{O} \left[ \frac{1}{N \ln(c-1)} \right].
\label{eq:PStail7}
\end{equation}

\noindent
In the infinite network limit one expects the probability $P( {\rm SPATH})$ to vanish.
This is due to the fact that as $N$ is increased (keeping $c$ constant) the typical distance between pairs of random
nodes increases. Thus the probability that an RW starting from a random initial node will follow the shortest path 
to a random target node decreases as $N$ is increased.

\begin{figure}
\centerline{
\includegraphics[width=6.5cm]{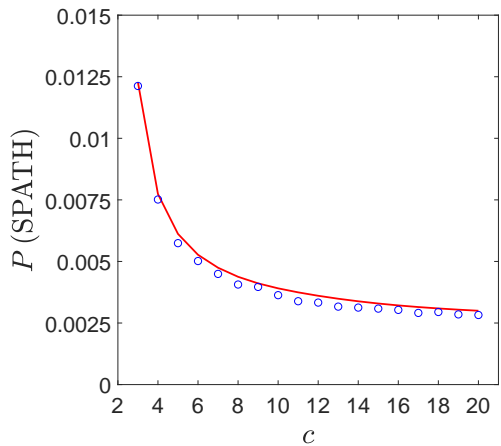}
}
\caption{
Analytical results (solid line) for the probability $P({\rm SPATH})$
that the first passage of an RW from a random initial node $i$ to a random
target node $j$ will occur via the ${\rm SPATH}$ scenario
on RRGs of size $N=1000$ as a function of the degree $c$.
The analytical results, obtained from Eq. (\ref{eq:PStail3}),
where $M_0$ is given by Eq. (\ref{eq:M_0a})
They are found to be in very good
agreement with the results obtained from computer simulations (circles).
}
\label{fig:7}
\end{figure}

In Fig. \ref{fig:7} we present
analytical results (solid line) for the probability $P({\rm SPATH})$
that the first-passage of an RW from a random initial node $i$ to a random
target node $j$ will occur via the ${\rm SPATH}$ scenario
on RRGs of size $N=1000$ as a function of the degree $c$.
The analytical results are obtained from Eq. (\ref{eq:PStail3}),
where $M_0$ is given by Eq. (\ref{eq:M_0a}).
They are found to be in very good
agreement with the results obtained from computer simulations.
The simulations are performed on an ensemble of RRGs that are 
constructed using the procedure presented in Sec. 2.
For each instance of a first-passage RW trajectory we select
a random initial node $i$ and a random target node $j$.
The trajectory of an RW that starts from node $i$ is recorded until
it visits node $j$ for the first time.
In order to qualify as an SPATH trajectory it should pass two tests.
First, it should not include any cycles. This means that the subgraph  
consisting of the nodes that the RW visited and the edges it used
along the way from $i$ to $j$ must be a tree.
Second, the distance between $i$ and $j$ on this subgraph must be
equal to the distance between $i$ and $j$ on the underlying RRG.
The fraction of first-passage RW trajectories that meet these two criteria 
is used to estimate $P({\rm SPATH})$.

The distribution of first-passage times from a random initial node $i$ to a random target node $j$,
under the condition that the RW follows the shortest path from $i$ to $j$,
is given by

\begin{equation}
P(T_{\rm FP}=t | {\rm SPATH}) = 
\sum_{\ell=1}^{t}
P(T_{\rm FP}=t|\ell,{\rm SPATH}) P(L=\ell | {\rm SPATH}),
\label{eq:PTfpts}
\end{equation}

\noindent
where

\begin{equation}
P(L=\ell | {\rm SPATH}) = 
\frac{ P({\rm SPATH} | \ell) P(L=\ell) }{P({\rm SPATH})}.
\label{eq:condSPATH}
\end{equation}

The probability 
$P(T_{\rm FP}=t|\ell,{\rm SPATH})$
makes a non-vanishing contribution to the sum only 
for even (and non-negative) values of $t-\ell$. 
Therefore, in Eq. (\ref{eq:PTfpts}) we need to distinguish between even
and odd values of $t$. 
In case that $t$ is even, the sum on the right hand
side is over even values of $\ell$, while
if $t$ is odd the sum is over
odd values of $\ell$.
Inserting 
$P(T_{\rm FP}=t|\ell,{\rm SPATH})$
from Eq. (\ref{eq:PTfp_ellS2}) into Eq. (\ref{eq:PTfpts}),
we find that for even values of the time $t$
 
\begin{eqnarray}
P    (T_{\rm FP} & = t    |   {\rm SPATH})  \simeq     
\frac{ (c-1)^{ \frac{t}{2} }  }{t c^t P({\rm SPATH})} \times
\nonumber \\
&    \sum\limits_{k=1}^{\frac{t}{2}}
 \frac{2k}{ (c-1)^{ k}  }
\binom{t}{\frac{t}{2} + k }
\left[ 1 + \frac{(c-1)^{2k}}{N} \right]
P(L=2k),
\label{eq:PTFPtSe}
\end{eqnarray}

\noindent
while for odd values of $t$

\begin{eqnarray}
P    (T_{\rm FP} & = t    |   {\rm SPATH})  \simeq     
\frac{(c-1)^{ \frac{t+1}{2} } }{t c^t P({\rm SPATH})} \times
\nonumber \\
&   \sum\limits_{k=1}^{\frac{t+1}{2}}
 \frac{2k-1}{ (c-1)^{ k}    }
\binom{t}{\frac{t-1}{2} + k}
\left[ 1 + \frac{(c-1)^{2k-1}}{N} \right]
P(L=2k-1).
\label{eq:PTFPtSo}
\end{eqnarray}

\noindent
The tail distribution of first-passage times in the SPATH scenario
is given by

\begin{equation}
P(T_{\rm FP}>t   |   {\rm SPATH}) =
\sum_{t'=t+1}^{\infty}
P(T_{\rm FP}=t'  |   {\rm SPATH}).
\label{eq:PTFPtail}
\end{equation}

\begin{figure}
\centerline{
\includegraphics[width=13cm]{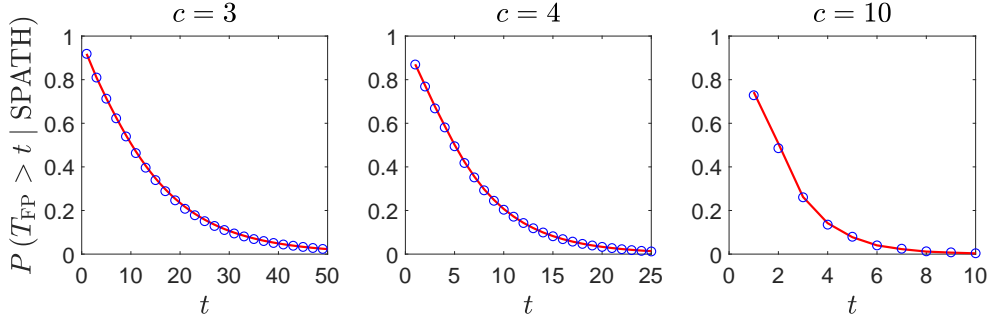}
}
\caption{
Analytical results (solid lines) for the tail distribution  
$P(T_{\rm FP} > t | {\rm SPATH})$ 
of first-passage times 
of RWs that follow the shortest
path from a random initial node $i$ 
to a random target node $j$
on an RRG
of size $N=1000$ 
and degrees
$c=3$ (left), $c=4$ (center)
and $c=10$ (right).
The analytical results,
obtained from Eqs. (\ref{eq:PTFPtail}),  
are in very good agreement with the results
obtained from computer simulations (circles).
}
\label{fig:8}
\end{figure}

In Fig. \ref{fig:8} we present
analytical results for the tail distribution  
$P(T_{\rm FP} > t | {\rm SPATH})$ 
of first-passage times (solid lines)
for RW trajectories that follow the shortest
path from a random initial node $i$ 
to a random target node $j$
on an RRG
of size $N=1000$ 
and degrees
$c=3$, $c=4$ 
and $c=10$.
The analytical results,
obtained from Eq. (\ref{eq:PTFPtail})  
are in very good agreement with the results
obtained from computer simulations (circles).

The first and second moments of the distribution of first-passage times 
between a random initial node $i$ and a random target node $j$,
under the condition that the RW follows the shortest path
between $i$ and $j$ are given by

\begin{equation}
\mathbb{E} [T_{\rm FP}^r | {\rm SPATH}] =
\sum_{\ell=1}^{N-1} 
\mathbb{E} [T_{\rm FP}^r | \ell,{\rm SPATH}] P(L=\ell | {\rm SPATH}),
\label{eq:ETfpr3}
\end{equation}

\noindent
where $r=1$ and $2$, respectively.
Inserting the first moment
$\mathbb{E} [T_{\rm FP} | \ell,{\rm SPATH}]$
from Eq. (\ref{eq:ETfp1p}) into Eq. (\ref{eq:ETfpr3})
and carrying out the summation, we obtain

\begin{equation}
\mathbb{E} [T_{\rm FP} | {\rm SPATH}] \simeq
\frac{1}{P({\rm SPATH})}
\left( \frac{c}{c-2} \right)
\sum_{\ell=1}^{N-1} \ell \left[ \left( \frac{1}{ c-1 } \right)^{\ell}  + \frac{1}{N} \right] P(L=\ell).
\end{equation}

\noindent
Using summation by parts and collecting terms, we obtain

\begin{equation}
\mathbb{E} [T_{\rm FP} | {\rm SPATH}] \simeq
\frac{1}{P({\rm SPATH})}
\left( \frac{c}{c-2} \right)
\left[  \frac{\langle L \rangle}{N} +
\frac{1}{ c-1 }
\sum_{\ell=0}^{N-1}   \frac{1 - (c-2) \ell}{(c-1)^{\ell}}  
P(L>\ell)   \right].
\end{equation}

\noindent
Using the results presented in Appendix B for the sums
$M_0$ and $M_1$, we obtain

\begin{equation}
\mathbb{E} [T_{\rm FP} | {\rm SPATH}]  \simeq 
\frac{1}{P({\rm SPATH})}
\left( \frac{c}{c-2} \right)
\left[
 \frac{\langle L \rangle}{N}
+ \frac{ M_0 - (c-2) M_1 }{ c-1 } \right],
\label{eq:ETFPS}
\end{equation}

\noindent
where $M_0$ is given by Eq. (\ref{eq:M_0a}), $M_1$ is given by
Eq. (\ref{eq:M_1a}) and $\langle L \rangle$ is the mean distance.
In the large network limit, one can express $M_0$ by Eq. (\ref{eq:M_0b}),
$M_1$ by Eq. (\ref{eq:M_1b}) and $P({\rm SPATH})$ by Eq. (\ref{eq:PStail6}).
This yields

\begin{equation}
\mathbb{E} [T_{\rm FP} | {\rm SPATH}]  \simeq 
\frac{1}{2} \left( \frac{c}{c-2} \right) 
\left[ \langle L \rangle + 1 - 4 \frac{c-1}{c(c-2)} + \frac{2}{\ln(c-1)} \right]
+ {\mathcal O} \left( \frac{1}{\ln N} \right).
\end{equation}
 
\noindent
The variance can be calculated by applying a similar approach to Eq. (\ref{eq:VTFP1}).
It yields

\begin{equation}
{\rm Var} [T_{\rm FP} | {\rm SPATH}]  \simeq 
\frac{1}{P({\rm SPATH})}
  \frac{4 c (c-1)}{(c-2)^3}  
\left[
 \frac{\langle L \rangle}{N}
+ \frac{ M_0 - (c-2) M_1 }{ c-1 } \right].
\label{eq:VTFPS}
\end{equation}
 
\noindent
Comparing Eqs. (\ref{eq:ETFPS}) and (\ref{eq:VTFPS}) it is found that 

\begin{equation}
{\rm Var} [T_{\rm FP} | {\rm SPATH}]  \simeq 
4 \frac{c-1}{(c-2)^2}
\mathbb{E} [T_{\rm FP} | {\rm SPATH}].
\end{equation}

\noindent
This implies that the distribution $P(T_{\rm FP}|{\rm SPATH})$ becomes narrower
as $c$ is increased.

\section{The distribution of first-passage times via non-SPATH trajectories}

In this section we derive a closed form expression for the distribution \\
$P(T_{\rm FP} > t |  \lnot {\rm SPATH})$ 
of first-passage times between a random initial node $i$ and a random target node $j$,
under the condition that the first-passage trajectory does not follow the shortest path
from $i$ to $j$.
In case that the distance $\ell$ between $i$ and $j$ is known,
the distribution of first-passage times, conditioned on $\ell$, is denoted by
$P(T_{\rm FP} > t |\ell,  \lnot {\rm SPATH})$.
Since an RW trajectory from $i$ to $j$ that does not follow the shortest
path must be longer than $\ell$, we conclude that
$P(T_{\rm FP} > t |\ell,  \lnot {\rm SPATH})=1$
for $t \le \ell$.

Starting from a random initial node $i$ 
the RW visits, on average, $\langle S \rangle_t$ distinct nodes 
up to time $t$.
In case that the distance between $i$ and $j$ is $\ell$,
the earliest time at which the RW may visit the target node
via a trajectory that does not follow the shortest path is $t=\ell+1$.
Therefore, the tail distribution of first-passage times,
conditioned on the distance $\ell$,
satisfies

\begin{equation}
P(T_{\rm FP}>t  |\ell, \lnot {\rm SPATH} ) = 1 - \frac{\langle S \rangle_{t-\ell}}{N},
\label{eq:P_FPT_tail}
\end{equation}

\noindent
where
\cite{Tishby2021b,Tishby2022b}

\begin{equation}
\langle S \rangle_t =   
  N \left[ 1 - e^{ -  \left( \frac{c-2}{c-1} \right) \frac{t}{N}   } \right].
\label{eq:Stlate2}
\end{equation}

\noindent
Inserting $\langle S \rangle_t$ from 
Eq. (\ref{eq:Stlate2}) into Eq. (\ref{eq:P_FPT_tail}), we obtain

\begin{equation}
P(T_{\rm FP} > t |\ell, \lnot {\rm SPATH}   )= 
\left\{
\begin{array}{ll}
1      & \ \ \ \  t \le \ell \\
\exp \left[ -    \left(  \frac{ c-2 }{c-1} \right) \frac{ t-\ell }{ N }         \right]     & \ \ \ \  t > \ell. 
\end{array}
\right.  
\label{eq:P_FPT_tail2}
\end{equation}

\noindent
The tail distribution of first-passage times is given by

\begin{equation}
P(T_{\rm FP} > t | \lnot {\rm SPATH}   ) =
\sum_{\ell=1}^{\infty}
P(T_{\rm FP} > t | \ell, \lnot {\rm SPATH}   )
P(L=\ell| \lnot {\rm SPATH}  ),
\label{eq:PTfps}
\end{equation}

\noindent
where

\begin{equation}
P(L=\ell | \lnot {\rm SPATH}) = 
\frac{ P(\lnot {\rm SPATH} | \ell) P(L=\ell) }{P(\lnot {\rm SPATH})}.
\label{eq:condSPATH2}
\end{equation}

\noindent
In Eq. (\ref{eq:condSPATH2})
the probability
$P(\lnot {\rm SPATH} | \ell)$
is given by Eq. (\ref{eq:PnoSell2})
and the probability
$P(\lnot {\rm SPATH}) = 1 - P({\rm SPATH})$,
where $P({\rm SPATH})$
is given by Eq. (\ref{eq:PStail3}).
Inserting 
$P(T_{\rm FP} > t |\ell, \lnot {\rm SPATH}   )$
from Eq. (\ref{eq:P_FPT_tail2}) into Eq. (\ref{eq:PTfps}),
we obtain

\begin{eqnarray}
& P(   T_{\rm FP} > t    |  \lnot {\rm SPATH}   )   = 
\sum_{\ell=t+1}^{\infty} P(L=\ell| \lnot {\rm SPATH})  
\nonumber  \\
 & + \exp \left[ -      \left( \frac{c-2}{c-1} \right)  \frac{t}{N}         \right]   
  \sum_{\ell=1}^{t} \exp \left[   \left( \frac{c-2}{c-1} \right)  \frac{\ell}{N}   \right] P(L=\ell |\lnot {\rm SPATH} ).     
\label{eq:PTfps2}
\end{eqnarray}

\noindent
Since the diameter of an RRG scales logarithmically with the network size
\cite{Bollobas1982},
in the large network limit and
for $t \gg \ln N$ 
Eq. (\ref{eq:PTfps2}) can be simplified to the form

\begin{equation}
P(T_{\rm FP} > t | \lnot {\rm SPATH}   )  =
\exp \left[ -   \left( \frac{c-2}{c-1} \right)  \frac{t}{N}     \right].
\label{eq:PTfps3}
\end{equation}

\noindent
The conditional probability mass function of first-passage times
for trajectories that do not follow the shortest path is given by

\begin{eqnarray}
P(T_{\rm FP}=t | \lnot {\rm SPATH}  ) &=& 
P(T_{\rm FP}>t-1| \lnot {\rm SPATH}  ) 
\nonumber \\
&-& P(T_{\rm FP}>t| \lnot {\rm SPATH}).
\label{eq:PTFPTSP}
\end{eqnarray}

\noindent
Inserting $P(T_{\rm FP}>t| \lnot {\rm SPATH})$ from Eq. (\ref{eq:PTfps3})
into Eq. (\ref{eq:PTFPTSP})
we obtain

\begin{equation}
P(T_{\rm FP} = t| \lnot {\rm SPATH})= 
\left\{ \exp \left[ {  \left( \frac{c-2}{c-1} \right)  \frac{1}{N}  } \right] - 1 \right\}
\exp \left[ -   \left( \frac{c-2}{c-1} \right)  \frac{t}{N}     \right].
\label{eq:P_FPT_pdf}
\end{equation}

\noindent
In the large network limit, where $N \gg 1$,
Eq. (\ref{eq:P_FPT_pdf}) can be approximated by

\begin{equation}
P(T_{\rm FP} = t| \lnot {\rm SPATH})= 
\left( \frac{c-2}{c-1} \right)  \frac{1}{N} 
\exp \left[ -   \left( \frac{c-2}{c-1} \right)  \frac{t}{N}    \right].
\label{eq:P_FPT_pdf2}
\end{equation}

The moments 
$\mathbb{E}[ T_{\rm FP}^r | \lnot {\rm SPATH}]$, $r=1,2,\dots$,
can be obtained 
from the tail-sum formula
\cite{Pitman1993}

\begin{equation}
\mathbb{E}[ T_{\rm FP}^r | \lnot {\rm SPATH}] = \sum_{t=0}^{\infty} 
[ (t+1)^r - t^r ] P(T_{\rm FP}>t| \lnot {\rm SPATH}).
\label{eq:FP_r}
\end{equation} 

\noindent
In particular, the
mean first-passage time 
$\mathbb{E}[ T_{\rm FP} | \lnot {\rm SPATH}]$ 
can be obtained by inserting $r=1$ in Eq. (\ref{eq:FP_r}),
which yields

\begin{equation}
\mathbb{E}[ T_{\rm FP} | \lnot {\rm SPATH}] = \sum_{t=0}^{\infty} P(T_{\rm FP}>t|  \lnot {\rm SPATH}    ).
\label{eq:FP_sum}
\end{equation}

\noindent
Similarly, inserting $r=2$ in Eq. (\ref{eq:FP_r}) we obtain the
second moment
$\mathbb{E}[ T_{\rm FP}^2 | \lnot {\rm SPATH}]$ 
of the distribution of first-passage times,
which is given by

\begin{equation}
\mathbb{E}[ T_{\rm FP}^2 | \lnot {\rm SPATH}]  = \sum_{t=0}^{\infty} 
(2 t + 1) P(T_{\rm FP}>t  |\lnot {\rm SPATH} ).
\label{eq:FP_sum2}
\end{equation} 

Inserting the tail distribution of 
Eq. (\ref{eq:PTfps3})
into Eq. (\ref{eq:FP_sum}) 
we obtain the mean first-passage time,
which is given by

\begin{equation}
\mathbb{E}[ T_{\rm FP}  | \lnot {\rm SPATH}]
= 
\frac{1}{1 - \exp \left[-   \left( \frac{c-2}{c-1} \right)  \frac{1}{N}    \right]}.
\label{eq:t_FPT_0}
\end{equation}

\noindent
In the large network limit, where $N \gg 1$,
Eq. (\ref{eq:t_FPT_0}) can be approximated
by

\begin{equation}
\mathbb{E}[ T_{\rm FP}  | \lnot {\rm SPATH}]  
\simeq 
\frac{c-1}{c-2}N,
\label{eq:t_FPT}
\end{equation}

\noindent
where $c \ge 3$.
The result of Eq. (\ref{eq:t_FPT}) was obtained in Ref.
\cite{Martin2010} 
using different considerations.
Interestingly, 
in the dense network limit of $c \gg 1$
it is found that
$\langle T_{\rm FP} \rangle \simeq  N$,
while in the dilute network limit of $c=3$,
$\langle T_{\rm FP} \rangle \simeq 2 N$.
This reflects the larger fraction of backtracking steps in 
RRGs of low degree.

Inserting the tail distribution of 
Eq. (\ref{eq:PTfps3})
into Eq. (\ref{eq:FP_sum2})
we obtain 
the second moment  

\begin{equation}
\mathbb{E}[ T_{\rm FP}^2  | \lnot {\rm SPATH}]
= 
\frac{ 1 + \exp \left[-   \left( \frac{c-2}{c-1} \right)  \frac{1}{N}   \right]  }
{ \left\{ 1 - \exp \left[-   \left( \frac{c-2}{c-1} \right)  \frac{1}{N}  \right] \right\}^2 }.
\label{eq:t_FPT_2}
\end{equation}

\noindent
Combining the results for the first and second moments,
we obtain the variance, which is given by

\begin{equation}
{\rm Var}[T_{\rm FP} | \lnot {\rm SPATH}] =  
\frac{ \exp \left[-\frac{c-2}{(c-1)N} \right] }{ \left\{ 1 - \exp \left[-\frac{c-2}{(c-1)N} \right] \right\}^2 }.
\end{equation}

\section{The overall distribution of first-passage times}

For any value of the distance $\ell$ between the initial node $i$ and the target
node $j$, the conditional distribution of first-passage times can be written in the form

\begin{eqnarray}
P(T_{\rm FP} = t| \ell) &=&
P(T_{\rm FP} = t| \ell, {\rm SPATH}) P( {\rm SPATH} | \ell) 
\nonumber \\
&+&
P(T_{\rm FP} = t| \ell, \lnot {\rm SPATH}) P( \lnot {\rm SPATH} | \ell ),
\label{eq:Ptfpt}
\end{eqnarray}

\noindent
where the conditional tail distributions
$P(T_{\rm FP} = t| \ell, {\rm SPATH})$
and
$P(T_{\rm FP} = t| \ell, \lnot {\rm SPATH})=P(T_{\rm FP} = t| \lnot {\rm SPATH})$
are given by Eqs. (\ref{eq:PTfp_ellS2}) and (\ref{eq:P_FPT_pdf2}),
respectively,
while the probabilities
$P( {\rm SPATH} | \ell)$
and
$P( \lnot {\rm SPATH} | \ell )$
are given by 
Eqs. (\ref{eq:PSell3}) and (\ref{eq:PnoSell2}),
respectively.

The mean first-passage time from a random initial node $i$ to a random target node $j$,
under the condition that the distance between
$i$ and $j$ is $\ell$,
is given by

\begin{eqnarray}
\mathbb{E}[T_{\rm FP}|\ell] &=&
\mathbb{E}[T_{\rm FP}| \ell, {\rm SPATH}] P({\rm SPATH}|\ell)
\nonumber \\
&+&
\mathbb{E}[T_{\rm FP}| \ell, \lnot {\rm SPATH}] P(\lnot {\rm SPATH}|\ell),
\label{eq:ETfpell}
\end{eqnarray}

\noindent
where 
$\mathbb{E}[T_{\rm FP}| \ell, {\rm SPATH}]$
and
$\mathbb{E}[T_{\rm FP}| \ell, \lnot {\rm SPATH}] = \mathbb{E}[T_{\rm FP}| \lnot {\rm SPATH}]$
are given by Eqs. (\ref{eq:ETfp1p}) and (\ref{eq:t_FPT_0}), respectively.

The overall distribution of first-passage times can be written in the form

\begin{eqnarray}
P(T_{\rm FP} = t) &=&
P(T_{\rm FP} = t| {\rm SPATH}) P( {\rm SPATH} ) 
\nonumber \\
&+&
P(T_{\rm FP} = t| \lnot {\rm SPATH}) P( \lnot {\rm SPATH} ),
\label{eq:Ptfpt}
\end{eqnarray}

\noindent
where 
$P(T_{\rm FP} = t| {\rm SPATH})$
is given by 
Eqs. (\ref{eq:PTFPtSe})  and  (\ref{eq:PTFPtSo}),
for even and odd times, respectively,
while
$P(T_{\rm FP} = t| \lnot {\rm SPATH})$
is given by Eq. (\ref{eq:P_FPT_pdf}).
The corresponding tail distribution is given by

\begin{equation}
P(T_{\rm FP}>t) = \sum_{t'=t+1}^{\infty} P(T_{\rm FP}=t'),
\label{eq:FPTail}
\end{equation}

\noindent
where $P(T_{\rm FP}=t)$ is given by Eq. (\ref{eq:Ptfpt}).

The mean first-passage time
$\langle T_{\rm FP} \rangle$
can be expressed in the form

\begin{eqnarray}
\langle T_{\rm FP} \rangle &=&
\mathbb{E}[T_{\rm FP}  | {\rm SPATH}] P( {\rm SPATH} ) 
\nonumber \\
&+&
\mathbb{E}[T_{\rm FP} | \lnot {\rm SPATH}] P( \lnot {\rm SPATH} ),
\label{eq:TFP}
\end{eqnarray}

\noindent
where
$\mathbb{E}[T_{\rm FP}  | {\rm SPATH}]$
is given by Eq. (\ref{eq:ETFPS})
and
$\mathbb{E}[T_{\rm FP} | \lnot {\rm SPATH}]$
is given by Eq. (\ref{eq:t_FPT_0}).
The mean first passage time
$\langle T_{\rm FP} \rangle$
was recently calculated for specific network instances and
a given pair of initial and target nodes.
\cite{Pitman2018,Bartolucci2021,Forster2022a,Forster2022b}.
These calculations were done using matrix methods that apply to
a wide range of network structures inclusing weighted networks.

\begin{figure}
\centerline{
\includegraphics[width=13cm]{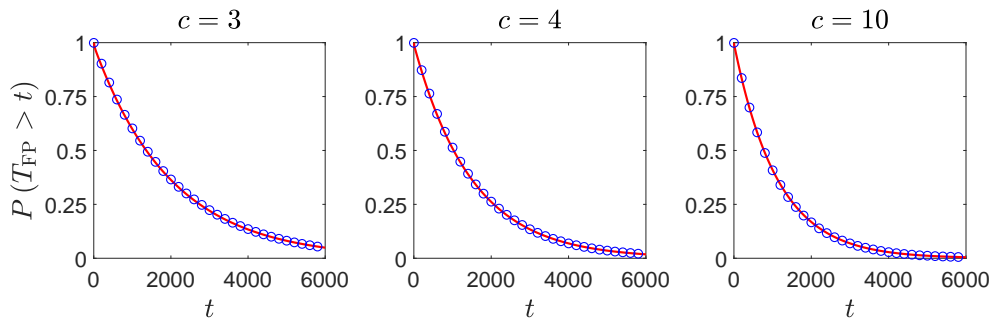}
}
\caption{
Analytical results (solid lines) for the 
tail distribution  
$P(T_{\rm FP} > t)$ 
of first-passage times 
of RWs on RRGs
of size $N=1000$ 
and degrees
$c=3$ (left), $c=4$ (center)
and $c=10$ (right).
The analytical results,
obtained from Eqs. (\ref{eq:FPTail})  
are in very good agreement with the results
obtained from computer simulations (circles).
}
\label{fig:9}
\end{figure}

In Fig. \ref{fig:9} 
we present analytical results for the tail distribution  
$P(T_{\rm FP}>t)$
of first-passage times
of RWs on RRGs 
of size $N=1000$ and 
$c=3$ (left), $c=4$ (center)
and $c=10$ (right).
The analytical results,
obtained from Eq. (\ref{eq:FPTail})  
are in very good agreement with the results
obtained from computer simulations (circles).
At the time scales shown here the first passage process is
dominated by the non-SPATH scenario.
Thus, the tail distribution of first passage times is essentially an exponential distribution.

\begin{figure}
\centerline{
\includegraphics[width=7cm]{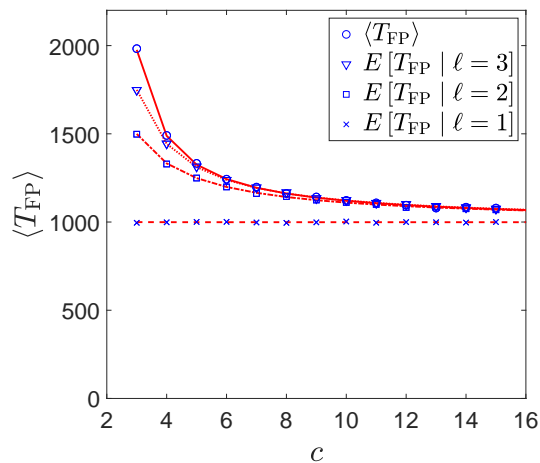}
}
\caption{
Analytical results (solid line) for the
mean first-passage time 
$\langle T_{\rm FP} \rangle$ 
of RWs on RRGs of size $N=1000$
as a function of the degree $c$. 
The analytical results,
obtained from Eq.
(\ref{eq:TFP}) 
are in very good agreement with 
the results obtained from computer simulations (circles).
We also present analytical results for the conditional expectation values
${\mathbb E}[T_{\rm FP}|L=\ell]$
for $\ell=1$ (dashed line), $\ell=2$ (dashed-dotted line) and $\ell=3$ (dotted line).
The analytical results, obtained from Eq. (\ref{eq:ETfpell}), are in very good
agreement with the simulation results ($\times$, {\scriptsize $\square$} and $\triangledown$, respectively).
As $\ell$ is increased the conditional expectation values converge towards
$\langle T_{\rm FP} \rangle$.
}
\label{fig:10}
\end{figure}

In Fig. \ref{fig:10} 
we present analytical results (solid line) for the
mean first-passage time 
$\langle T_{\rm FP} \rangle$ 
as a function of the degree 
$c$ for RWs on RRGs 
of size $N=1000$.
The analytical results,
obtained from Eq.
(\ref{eq:TFP}) 
are in very good agreement with 
the results obtained from computer simulations (circles).
The mean first passage time
is a monotonically decreasing function of $c$, that 
converges towards 
$\langle T_{\rm FP} \rangle \simeq N$
in the dense-network limit.
We also present analytical results for the conditional expectation values
${\mathbb E}[T_{\rm FP}|L=\ell]$
for $\ell=1$ (dashed line), $\ell=2$ (dashed-dotted line) and $\ell=3$ (dotted line).
The analytical results, obtained from Eq. (\ref{eq:ETfpell}), are in very good
agreement with the simulation results (symbols).
In the special case of $\ell=1$ the conditional expectation value
${\mathbb E}[T_{\rm FP}|L=1]$ is essentially independent of $c$, 
resulting in a plateau in Fig. \ref{fig:10}.
More precisely, it satisfies 
${\mathbb E}[T_{\rm FP}|L=1]=N - 1 + \mathcal{O}(1/c)$.
This result is consistent with the result for the mean first return time,
$\langle T_{\rm FR} \rangle = N$
\cite{Tishby2021b}.
The difference of one step is due to the fact that the first step in the
first return problem brings the RW to a distance $\ell=1$ from the
initial node (which is also the target node).
This implies that
$\langle T_{\rm FR} \rangle = {\mathbb E}[T_{\rm FP}|L=1] + 1$.
As $\ell$ is increased the conditional expectation values 
${\mathbb E}[T_{\rm FP}|L=\ell]$
converge towards
$\langle T_{\rm FP} \rangle$.
This convergence exemplifies the fact that the SPATH scenario is significant
only in the sparse network limit where $c$ is small and
only when the distance $\ell$ between the initial node and target node
is very small.

\section{Summary and Discussion}

We presented analytical results for the distribution of
first-passage times of
RWs on RRGs of 
degree $c \ge 3$ and of a finite size $N$.
We calculated
the tail distribution of first-passage times
$P( T_{\rm FP} > t )$
from a random initial node $i$ to a random target node $j$.
We identified two types of first-passage trajectories,
namely trajectories that follow the shortest path between 
$i$ and $j$ (SPATH trajectories) and those that deviate from the shortest path
(non-SPATH trajectories).
Using this distinction we derived closed form analytical expressions
for the tail distribution of first-passage times for each 
one of the two scenarios.
It was found that the probability $P({\rm SPATH}|\ell)$ that the
first passage trajectory from an initial node $i$ to a target node $j$ that resides 
at a distance $\ell$ from $i$ will be an SPATH trajectory decays exponentially 
like $1/(c-1)^{\ell}$ with the distance $\ell$.
Therefore, the SPATH scenario is probable only when 
the network is sparse and the distance $\ell$
between the initial node and the target node is small.
The probability that the first passage trajectory from a random initial node $i$ to 
a random target node $j$ will be an SPATH trajectory scales like
$P({\rm SPATH}) \sim \ln N/N$, which decreases as the network size is increased. 
Thus, overall the SPATH scenario is a low probability scenario.
In the SPATH scenario, the mean first passage time scales like
$\mathbb{E}[T_{\rm FP}|{\rm SPATH}] \sim \langle L \rangle \sim \ln N$,
where $\langle L \rangle$ is the mean distance between pairs of nodes in the network.
In the non-SPATH scenario, the mean first passage time scales like
$\mathbb{E}[T_{\rm FP}| \lnot {\rm SPATH}] \sim N$.

The first passage process studied in this paper is an important landmark in the life-cycle of
an RW on an RRG.   
Another important event is the first hitting process, which is the first time
in which the RW enters a previously visited node 
\cite{Tishby2017,Tishby2021a}.
This event represents the interaction of an RW with its own trajectory
rather than with a pre-defined target node.
The characteristic time scale of the first hitting process is 
$t \sim \min \{c,\sqrt{N} \}$,
namely $t \sim c$ in dilute networks and 
$t \sim \sqrt{N}$ in dense networks.
In both cases the first hitting time is typically much shorter than the first-passage time
\cite{Tishby2017,Tishby2021a}.
Note that the terminology on first hitting processes in not universal across the
scientific literature, and in some cases first passage times are referred to as
first hitting times.

Yet another important event,
which occurs at much longer time scales,
is the step at which the RW completes 
visiting all the nodes in the network.
The time at which this happens is called the cover-time, which scales
like $t \sim N \ln N$
\cite{Tishby2022b,Cooper2005}.
This means that on average an RW visits each node $\ln N$ times before
it completes visiting all the nodes in the network at least once.
The distribution $P(T_{\rm C} \le t)$ of cover times can be obtained from
the distribution of first passage times
using the framework of extreme-value theory.
The cover time 
can be considered as the maximum among 
$N-1$ first passage times from the initial node $i$
to all the other nodes in the network.
Therefore, under the assumption that the distributions of first passage times
for different nodes are independent, the distribution of cover times satisfies

\begin{equation}
P(T_{\rm C} > t) = 1 - [ 1 - P(T_{\rm FP} > t) ]^{N-1}.
\label{eq:CoverFP}
\end{equation}

\noindent


\noindent
The cover time involves very long trajectories, where the non-SPATH
scenario is dominant over the SPATH scenario.
Thus, in this context it is suitable to approximate the distribution of 
first passage times $P(T_{\rm FP} > t)$ by 
$P(T_{\rm FP} > t| \lnot {\rm SPATH})$,
which is given by Eq. (\ref{eq:PTfps3}).
Inserting this expression into Eq. (\ref{eq:CoverFP}), 
it is found that the distribution of cover times follows a discrete Gumbel distribution,
known from extreme value theory,
which takes the form
\cite{Gumbel1935}

\begin{equation}
P(T_{\rm C} > t) \simeq
1 - \exp \left[ -   \exp \left( {- \frac{ t - \mu } {\beta } } \right) \right],
\label{eq:Gumbel3}
\end{equation}

\noindent
where

\begin{equation}
\mu =  \frac{c-1}{c-2}  N \ln (N-1)
\end{equation}

\noindent
is called the location parameter and

\begin{equation}
\beta = \frac{c-1}{c-2} N 
\end{equation}

\noindent
is called the scale parameter.
This is in agreement with the results of Ref. \cite{Tishby2022b}.
The location parameter $\mu$ is equal to the mode of the Gumbel distribution.
The scale parameter $\beta$ is equal to the standard deviation up to a constant
factor of order $1$.
The Gumbel distribution often emerges as the distribution of the 
maxima among sets of $n$ independent random variables drawn from the same distribution.
It is one of the three possible families of extreme value distributions
specified by the extreme value theory,
namely the Gumbel, Fr\'echet and Weibull families
\cite{Gumbel1935,Frechet1927,Fisher1928,Mises1936,Gnedenko1943}.
The distribution of first passage times exhibits an exponential tail.
It thus meets the criterion for the emergence of the Gumbel distribution
in the Fisher-Tippet-Gnedenko theorem
\cite{Frechet1927,Fisher1928,Mises1936,Gnedenko1943}.
Note that the first passage times of adjacent target nodes may be correlated.
However, such correlations appear to have little effect on the distribution of
cover times.


This work was supported by the Israel Science Foundation grant no. 
1682/18.

\appendix

\section{Calculation of the combinatorial factor $B(t,\ell)$}

In this Appendix we solve the recursion equations for the number $B(t,\ell)$ of distinct
RW trajectories of length $t$ between a pair of nodes $i$ and $j$ which reside at a distance
$\ell$ apart from each other and reach $j$ for the first time after $t$ time steps.
The recursion equations are shown in Eq. (\ref{eq:Atell}).
To solve the equation we express $B(t,\ell)$ in the form

\begin{equation}
B(t,0) = c(c-1)^{\frac{t-2}{2}} b(t,0)
\label{eq:A1}
\end{equation}

\noindent
for $\ell=0$, and in the form

\begin{equation}
B(t,\ell) =  (c-1)^{\frac{t-\ell}{2}} b(t,\ell)
\label{eq:A2}
\end{equation}

\noindent
for $\ell \ge 1$.
Inserting these relations into Eq. (\ref{eq:Atell}), we obtain

\begin{equation}
b(t,\ell) = b(t-1,\ell-1) + b(t-1,\ell+1),
\label{eq:btl}
\end{equation}

\noindent
for $\ell \ge 2$ and $t \ge \ell$.
In the special cases of $\ell=0,1$ and $t \ge 2$, we obtain
$b(t,\ell)=b(t-1,\ell+1)$.
The transformation presented in Eqs. (\ref{eq:A1}) and (\ref{eq:A2})
factors out the dependence on the mean degree $c$ and 
greatly simplifies the recursion equations for $b(t,\ell)$.
In order to solve the recursion equations given by Eq. (\ref{eq:btl}),
we introduce the generating function

\begin{equation}
G_t(x) = \sum_{\ell = 0}^{\infty} b(t,\ell) x^{\ell}.
\label{eq:Ftx0}
\end{equation}

\noindent
Multiplying Eq. (\ref{eq:btl}) by $x^{\ell}$ and summing up over $\ell$,
we obtain

\begin{equation}
G_t(x) = \left( x + \frac{1}{x} \right) \left[ G_{t-1}(x) - G_{t-1}(0) \right],
\label{eq:Ftx}
\end{equation}

\noindent
for $t \ge 2$,
while for $t=1$, $G_1(x)=x$.
Applying the Z-transform with respect to time, we obtain
another generating function of the form

\begin{equation}
\Phi(x,\omega) = \sum_{t=1}^{\infty}
G_t(x) \omega^t.
\label{eq:Phixo1}
\end{equation}

\noindent
Multiplying Eq. (\ref{eq:Ftx}) by $\omega^t$ and summing up over $t$,
we obtain

\begin{equation}
\Phi(x,\omega) = \omega x +  \omega \left( x + \frac{1}{x} \right) 
\left[ \Phi(x,\omega) - \Phi(0,\omega) \right].
\end{equation}

\noindent
Solving for $\Phi(x,\omega)$, we obtain

\begin{equation}
\Phi(x,\omega) =
\left[ 1 - \left( 1 - \frac{x^2 + 1}{x} \omega \right)^{-1} \right]
\left[ \Phi(0,\omega) - \frac{x^2}{x^2+1} \right],
\label{eq:Pxom}
\end{equation}

\noindent
where $\Phi(0,\omega)$ is a boundary condition.
In fact, one can determine $\Phi(0,\omega)$ directly from combinatorial considerations.
Note that

\begin{equation}
\Phi(0,\omega) = \sum_{t=1}^{\infty}
b(t,0) \omega^t.
\end{equation}

\noindent
As mentioned in the main text, the
coefficient $B(t,0)$
accounts for the number of distinct RW trajectories that return
to the initial node after $t$ time under the condition that these trajectories 
have not visited the initial node at earlier times
\cite{Tishby2021b}.
It can be expressed in terms
of the Catalan numbers $C_k$, which count the number of mountain ranges
of length $2k$
\cite{Audibert2010}.
More specifically, 

\begin{equation}
B(t,0) = 
\left\{
\begin{array}{ll}
c (c-1)^{ \frac{t}{2} - 1 } C_{ \frac{t}{2} - 1 }  & \ \ \ \ \ \ \  t   \ \   {\rm even} \\
0 & \ \ \ \ \ \ \  t   \ \  {\rm odd},
\end{array}
\right.  
\label{eq:Bt0}
\end{equation}

\noindent
where

\begin{equation}
C_{ k } = \frac{ (2k)! }{k! \left( k+1 \right)!  } 
= \frac{1}{2k+1} \binom{2k+1}{k+1},
\label{eq:Catalan}
\end{equation}

\noindent
and $\binom{n}{m}$ is the binomial coefficient.

From the definition of $b(t,0)$ and Eq. (\ref{eq:Bt0}),
it is found that for
even time steps
$b(t,0)$ coincides with the Catalan number $C_{ \frac{t-2}{2} }$,
while for odd time steps it vanishes.
Therefore, 

\begin{equation}
\Phi(0,\omega) = 
\sum_{k=0}^{\infty}
C_k \omega^{2k+2}.
\label{eq:P0om}
\end{equation}

\noindent
Inserting $\Phi(0,\omega)$ from Eq. (\ref{eq:P0om}) into Eq. (\ref{eq:Pxom})
and expanding it in powers of $\omega$, we obtain

\begin{eqnarray}
\Phi(x,\omega) &=&
x \omega + (x^2 + 1)\omega^2
\nonumber \\
&+& \sum_{t=3}^{\infty}
\left[  x \left( \frac{x^2+1}{x} \right)^{t-1}
-
\sum_{k=0}^{ \lfloor \frac{t-3}{2} \rfloor }
C_k \left( \frac{x^2+1}{x} \right)^{t-2k-2} 
\right]
\omega^t,
\end{eqnarray}

\noindent
where the floor function $\lfloor x \rfloor$ is the largest integer which is smaller than $x$.
Following Eq. (\ref{eq:Phixo1}) one can extract the generating function $G_t(x)$
from the last equation.
For $t=2$ we obtain $G_2(x)=x^2+1$
while for $t \ge 3$ we obtain

\begin{equation}
G_t(x) =
x \left( \frac{x^2+1}{x} \right)^{t-1}
-
\sum_{k=0}^{ \lfloor \frac{t-3}{2} \rfloor }
C_k \left( \frac{x^2+1}{x} \right)^{t-2k-2}.
\end{equation}

\noindent
Summarizing these results, 
we obtain

\begin{equation}
G_t(x) =
\left\{
\begin{array}{ll}
\frac{2}{t}
\binom{t-2}{ \frac{t-2}{2} } 
\, _2F_1 \left[ \left.
\begin{array}{c}
1,\frac{t-1}{2} \\
\frac{t+2}{2}
\end{array}
\right| \frac{4x^2}{(x^2+1)^2}
\right]  & \ \ \ \ \ \ \  t  \ \   {\rm even} \\
\frac{2}{t+1}
\binom{t-1}{ \frac{t-1}{2} } 
\frac{x}{x^2+1}
\, _2F_1 \left[ \left.
\begin{array}{c}
1, \frac{t}{2} \\
\frac{t+3}{2}
\end{array}
\right| \frac{4x^2}{(x^2+1)^2}
\right]  & \ \ \ \ \ \ \  t   \ \ \  {\rm odd}.
\end{array}
\right.  
\end{equation}

\noindent
For the analysis below we use the identity
(Section 2.11, Eq. (11) in Ref.
\cite{Erdelyi1953})

\begin{eqnarray}
\, _2F_1 \left( \left.
\begin{array}{c}
a,b \\
a+b+\frac{1}{2}
\end{array}
\right| z
\right)
& =
\left( \frac{2}{\sqrt{1-z}+1} \right)^{2 a}
\times
\nonumber \\
& \, _2F_1 \left( \left.
\begin{array}{c}
2a,a-b+\frac{1}{2} \\
a+b+\frac{1}{2}
\end{array}
\right| \frac{ \sqrt{1-z}-1 }{\sqrt{1-z} + 1}
\right).
\label{eq:I}
\end{eqnarray}

\noindent
Inserting $a=1$ and $b=k+1/2$ into Eq. (\ref{eq:I}),
we obtain

\begin{equation}
\, _2F_1 \left[ \left.
\begin{array}{c}
1,k+\frac{1}{2} \\
k+2
\end{array}
\right| \frac{4x^2}{(x^2+1)^2}
\right]
=
\left( x^2 + 1 \right)^{2}
\, _2F_1 \left( \left.
\begin{array}{c}
2,1-k \\
k+2
\end{array}
\right| -x^2
\right).
\label{eq:J}
\end{equation}

\noindent
Using the definition of the hypergeometric function on the right hand side, we obtain

\begin{equation}
\, _2F_1 \left[ \left.
\begin{array}{c}
1,k+ \frac{1}{2} \\
k+2
\end{array}
\right| \frac{4x^2}{(x^2+1)^2}
\right]
=
\left( x^2 + 1 \right)^{2}
\sum_{m=0}^{\infty}
\left[ \frac{ (2)_m(1-k)_m }{ (k+2)_m } \right]
\left[ \frac{ (-x^2)^m }{m!} \right],
\label{eq:J2}
\end{equation}

\noindent
where $(a)_m$ is the (rising) Pochhammer symbol
\cite{Olver2010}.
For integer values $a>0$ it can be expressed as

\begin{equation}
(a)_m = \frac{ (a+m-1)! }{ (a-1)! },
\end{equation}

\noindent
while for negative integer values $a<0$  

\begin{equation}
(a)_m = 
\left\{
\begin{array}{ll}
(-1)^m \frac{ (-a)! }{ (-a-m)! } & m \le -a \\
0 & m > -a
\end{array}
\right.
\end{equation}

\noindent
Inserting these expressions into Eq. (\ref{eq:J2}), we obtain

\begin{equation}
\, _2F_1 \left[ \left.
\begin{array}{c}
1,k+\frac{1}{2} \\
k+2
\end{array}
\right| \frac{4x^2}{(x^2+1)^2}
\right]
=
\left( x^2 + 1 \right)^{2}
\sum_{m=0}^{k-1}
\frac{ \binom{k-1}{m} }{ \binom{ k+1+m }{ m } }
(m+1) x^{2m}.
\label{eq:J3}
\end{equation}

\noindent
Therefore,

\begin{equation}
G_t(x) =
\left\{
\begin{array}{ll}
\frac{1}{t-1}
\binom{t-1}{\frac{t}{2}}
+
\sum\limits_{m=1}^{\frac{t-2}{2}}
\frac{ 2m  }{t}
\binom{t}{\frac{t}{2}+m} x^{2m}
+ x^t 
&   \ \ \ \ \ \   t   \ \    {\rm even} \\
\sum\limits_{m=0}^{\frac{t-3}{2}}
\frac{ 2m+1  }{t}
\binom{t}{  \frac{t+1}{2}+m  } x^{2m+1}
+ x^t   
& \ \ \ \ \ \  t   \ \ \  {\rm odd}.
\end{array}
\right.  
\end{equation}

\noindent
Going back to the definition of the generating function $G_t(x)$
in Eq. (\ref{eq:Ftx0}) one can identify the coefficients $b(t,\ell)$.
It turns out that the results for both even and odd values of $t$ can 
be expressed by the same expression, which takes the form

\begin{equation}
b(t,\ell) = \frac{\ell}{t} \binom{t}{ \frac{t+\ell}{2} },
\label{eq:btell}
\end{equation}

\noindent
under the condition that $t-\ell$ is even.
Interestingly, for $\ell > 0$ these combinatorial factors coincide
with the number of RW trajectories in a semi-infinite one dimensional system
that start from $x=0$ and reach the site $x=\ell$ for the first time at time $t$
\cite{Redner2001}.
Inserting $b(t,\ell)$ from Eq. (\ref{eq:btell}) into Eqs. (\ref{eq:A1}) and (\ref{eq:A2}) 
it is found that

\begin{equation}
B(t,\ell) =  
 \left\{
\begin{array}{ll}
\frac{\ell}{t} 
\binom{t}{   \frac{t+\ell}{2}   } 
(c-1)^{ \frac{t-\ell}{2} }  & \ \ \ \ \ \ \  t-\ell  \ \   {\rm even} \\
0 & \ \ \ \ \  \ \  t-\ell  \ \  {\rm odd}.
\end{array}
\right.  
\label{eq:A2b}
\end{equation}

\section{Useful sums involving $P(L>\ell)$}

The discrete Laplace transform of $P(L>\ell)$ is defined by

\begin{equation}
{\mathcal L}[P(L>\ell)](s) =
\sum_{\ell=0}^{\infty}
e^{- s \ell} P(L>\ell).
\label{eq:Laplace}
\end{equation}

\noindent
Carrying out the summation, it was found in Ref. 
\cite{Tishby2022} that

\begin{eqnarray}
    \mathcal{L} & \{ P(L>\ell) \}(s)  = 
\frac{ \exp \left[ \left( \frac{c}{c-2} \right) \frac{1}{N} \right] }{\ln (c-1) }
{\rm E}_{1 + \frac{s}{\ln (c-1)}} \left[ \left( \frac{c}{c-2} \right) \frac{1}{N} \right]
+ \frac{1}{2} 
\nonumber \\
&-  \left[ \left( \frac{c}{c-2} \right) \frac{1}{N} \right]
\left\{ \frac{1}{2} \coth \left[ \frac{s-\ln(c-1)}{2} \right] 
- \frac{1}{s-\ln(c-1)}  \right\}
\nonumber \\
&+ 
\left[ \frac{1}{2} \coth \left( \frac{s}{2} \right) - \frac{1}{s}  \right]
\left[ 1 + \left( \frac{c}{c-2} \right) \frac{1}{N} \right]
+  \mathcal{O} \left( \frac{1}{N^2} \right),
\label{eq:LapPell2}
\end{eqnarray}

\noindent
where ${\rm E}_m(x)$ is the generalized exponential integral
(Eq. 8.19.3 in Ref. \cite{Olver2010}).

The Laplace transform is a useful tool for the evaluation of various
sums that include $P(L>\ell)$. 
In particular, the mean distance 
$\langle L \rangle$
between pairs of nodes in an RRG
can be expressed 
using the tail-sum formula \cite{Pitman1993}.
It takes the form

\begin{equation}
\langle L \rangle =
\sum_{\ell=0}^{\infty}
P(L>\ell) =
\mathcal{L} \{ P(L>\ell) \}(s=0).
\end{equation}

\noindent
Plugging in $s=0$ in Eq. (\ref{eq:LapPell2}), we obtain

\begin{eqnarray}
\langle L \rangle &=&  
\frac{ \exp \left[ \left( \frac{c}{c-2} \right) \frac{1}{N} \right] }{\ln (c-1)}
{\rm E_1} \left[ \left( \frac{c}{c-2} \right) \frac{1}{N} \right]
+ \frac{1}{2}
\nonumber \\
&+&
\frac{1}{2}
\left[ \left( \frac{c}{c-2} \right) 
\frac{1}{N} \right] 
\left\{ \coth \left[ \frac{\ln (c-1) }{2} \right] - \frac{2}{\ln (c-1)} \right\}
\nonumber \\
&+& \mathcal{O} \left(  \frac{1}{N^2}  \right).
\label{eq:EM5}
\end{eqnarray}

\noindent
In the large network limit, Eq. (\ref{eq:EM5}) can be reduced to

\begin{equation}
\langle L \rangle  =    
\frac{\ln N}{\ln (c-1)} 
+ \frac{1}{2}
- \frac{ \ln \left( \frac{c}{c-2} \right) + \gamma}{\ln (c-1)}  
+ \mathcal{O} \left(  \frac{\ln N}{N}  \right).
\label{eq:EM7}
\end{equation}

Below we evaluate several other sums that appear in the calculation
of the moments $\mathbb{E}[T_{\rm FP}^r|{\rm SPATH}]$ of the distribution
of first-passage times.
In particular, we consider the family of sums given by

\begin{equation}
M_n = \sum_{\ell=0}^{\infty}
\ell^n \left( \frac{1}{c-1} \right)^{\ell}  P(L>\ell),
\label{eq:S0}
\end{equation}

\noindent
where $P(L>\ell)$ is give by Eq. (\ref{eq:taildist}).
These sums can be expressed by

\begin{equation}
M_n = (-1)^n \frac{d^n}{ds^n} {\mathcal L}[P(L>\ell)](s) \bigg\vert_{s=\ln(c-1)}.
\label{eq:M_n}
\end{equation}

\noindent
Inserting $\mathcal{L} \{ P(L>\ell) \}(s)$ from Eq. (\ref{eq:LapPell2}) into Eq. (\ref{eq:M_n})
for $n=0$, we obtain

\begin{eqnarray}
M_0   =  \frac{c-1}{c-2} &+
   \frac{ \exp \left[ \left( \frac{c}{ c-2 } \right) \frac{1}{N} \right] {\rm E_1} \left[ - \left( \frac{c}{c-2} \right) \frac{1}{N} \right] -1 }{\ln(c-1)}
\left( \frac{c}{c-2} \right)  \frac{1}{N}  
\nonumber \\
 &+ \frac{1}{2}  \left( \frac{c}{c-2} \right)^2 \frac{1}{N}   +  {\mathcal O} \left( \frac{1}{N^2} \right).
\label{eq:M_0a}
\end{eqnarray}

\noindent
In the large network limit, Eq. (\ref{eq:M_0a}) can be reduced to

\begin{equation}
M_0 = \frac{c-1}{c-2} -
\frac{c}{c-2} \left[ \frac{ \ln N  }{\ln(c-1)} 
+ \frac{ \ln \left( \frac{c-2}{c} \right) - \gamma +1 }{\ln (c-1)}
- \frac{1}{2} \left( \frac{c}{ c-2 }  \right) \right] \frac{1}{N}
+ {\mathcal O} \left( \frac{1}{N^2} \right).
\label{eq:M_0b}
\end{equation}

\noindent
Inserting $\mathcal{L} \{ P(L>\ell) \}(s)$ from Eq. (\ref{eq:LapPell2}) into Eq. (\ref{eq:M_n})
for $n=1$, we obtain

\begin{eqnarray}
& M_1  =  
\frac{ c-1 }{(c-2)^2} - \frac{1}{[\ln(c-1)]^2}
\nonumber \\
&+ 
\left( \frac{c}{c-2} \right)
\left\{ \frac{c-1}{(c-2)^2} 
 +
\frac{ 
\exp \left[ \left( \frac{c}{ c-2 } \right) \frac{1}{N}   \right]  
G_{2,3}^{3,0}\left[ \left( \frac{c}{ c-2 } \right) \frac{1}{N} \left| 
\begin{array}{c}
1,1    \\
-1,0,0
\end{array}
\right. \right] - 1 }
{[\ln(c-1)]^2} 
\right\} \frac{1}{N}
\nonumber \\
&+  
{\mathcal O} \left( \frac{1}{N^2} \right),
\label{eq:M_1a}
\end{eqnarray}

\noindent
where $G(\cdot)$ is the Meijer function 
\cite{Olver2010}.
The Meijer function can be expressed in terms of more elementary functions,
in the form

\begin{eqnarray}
G_{2,3}^{3,0} & \left[ \left( \frac{c}{ c-2 } \right) \frac{1}{N} \left| 
\begin{array}{c}
1,1    \\
-1,0,0
\end{array}
\right. \right]  = 
\left( \frac{ c-2  }{c} \right) N \exp \left[ - \left( \frac{c}{ c-2 } \right) \frac{1}{N} \right]
\nonumber \\
&- \frac{1}{2} \left[ \ln \left( \frac{c-2}{c} N \right) \right]^2
+  \gamma \ln \left( \frac{c-2}{c} N \right)
- \frac{\pi^2}{12} - \frac{ \gamma^2 }{2}
\nonumber \\
&+  \left( \frac{c}{ c-2 } \right) \frac{1}{N}  
\ _3F_3 \left[ \left.
\begin{array}{c}
1,1,1 \\
2,2,2
\end{array}
\right| - \left( \frac{c}{ c-2 } \right) \frac{1}{N} 
\right]
\nonumber \\
&+  {\rm Chi} \left[ \left( \frac{c}{ c-2 } \right) \frac{1}{N} \right]
- {\rm Shi} \left[ \left( \frac{c}{ c-2 } \right) \frac{1}{N} \right],
\end{eqnarray}

\noindent
where ${\rm Chi}(x)$ and ${\rm Shi}(x)$ are the
hyperbolic cosine and sine integrals, respectively
\cite{Olver2010}.
In the large network limit, Eq. (\ref{eq:M_1a}) can be reduced to

\begin{eqnarray}
M_1 &=& 
\frac{ c-1 }{(c-2)^2} 
- \frac{c}{2(c-2) [\ln(c-1)]^2} \frac{  
\left[ \ln \left( \frac{c-2}{c} N \right) \right]^2 }{N}
\nonumber \\
&+& \frac{ (\gamma-1)c}{ (c-2) [\ln(c-1)]^2} \frac{ \ln \left( \frac{c-2}{c} N \right) }{N}
+ \frac{c(c-1)}{(c-2)^3} \frac{1}{N}
\nonumber \\
&-& \frac{ c \left[  \pi^2  +6  (\gamma-2)\gamma + 12 \right] }
{12(c-2) [\ln(c-1)]^2} \frac{1}{N}
 +  {\mathcal O} \left( \frac{1}{N^2} \right).
\label{eq:M_1b}
\end{eqnarray}

\noappendix



\section*{References}

\begin{thebibliography}{10}



\bibitem{Lawler2010b}
Lawler G F 2010
{\it Random Walk and the Heat Equation}
(Providence: American Mathematical Society)

\bibitem{Lawler2010a}
Lawler G F and Limic V 2010
{\it Random Walk: A Modern Introduction}
(Cambridge: Cambridge University Press)

\bibitem{ben-Avraham2000}
ben-Avraham D and Havlin S 2000 
{\it Diffusion and Reactions in Fractals and Disordered Systems} 
(Cambridge: Cambridge University Press)

\bibitem{Noh2004}
Noh D J and Rieger H 2004
Random walks on complex networks
{\it Phys. Rev. Lett.} {\bf 92} 118701 




\bibitem{Masuda2017}
Masuda N, Porter M A and Lambiotte R 2017
Random walks and diffusion on networks
{\it Physics Reports} {\bf 716} 1


\bibitem{Havlin2010}
Havlin S and Cohen R 2010
{\it Complex Networks: Structure, Robustness and Function}
(Cambridge University Press, New York).

\bibitem{Newman2010}
Newman M E J 2018 {\it Networks: an Introduction, Second Edition} 
(Oxford: Oxford University Press).


\bibitem{Pastor-Satorras2001}
Pastor-Satorras R and  Vespignani A 2001
Epidemic spreading in scale-free networks
{\it Phys. Rev. Lett.} {\bf 86} 3200 

\bibitem{Barrat2012}
Barrat A, Barth\'elemy M and Vespignani A 2012
{\it Dynamical Processes on Complex Networks}
(Boston: Cambridge University Press)



\bibitem{Debacco2015}
De Bacco C, Majumdar S N and Sollich P 2015 
The average number of distinct sites visited by a random walker on random graphs
{\it J. Phys. A} {\bf 48} 205004


\bibitem{Masuda2004}
Masuda N and Konno N 2004
Return times of random walk on generalized random graphs
{\it Phys. Rev. E} {\bf 69} 066113




\bibitem{Redner2001}
Redner S 2001 
{\it A Guide to First Passage Processes}
(Cambridge: Cambridge University Press)





\bibitem{Finch2003}
Finch S R 2003
{\it Mathematical Constants}, section 5.9
(Cambridge: Cambridge University Press)


\bibitem{Sood2005}
Sood V, Redner S and ben-Avraham D 2005
First-passage properties of the Erd{\H o}s–R\'enyi random graph
{\it J. Phys. A} {\bf 38} 109

\bibitem{Baronchelli2006}
Baronchelli A and Loreto V 2006
Ring structures and mean first passage time in networks
{\it Phys. Rev. E}{\bf 73} 026103



\bibitem{Lau2010}
Lau H W and Szeto K Y 2010
Asymptotic analysis of first passage time in complex networks
{\it EPL} {\bf 90} 40005 


\bibitem{Bartolucci2021}
Bartolucci S, Caccioli F, Caravelli F and Vivo P 2021
"Spectrally gapped" random walks on networks: 
a Mean First Passage Time formula
{\it SciPost Phys.} {\bf 11} 088 



\bibitem{Bassolas2021}
Bassolas A and Nicosia V 2021
First-passage times to quantify and compare structural correlations and
heterogeneity in complex systems 
{\it Communications Physics} {\bf 4} 1 


\bibitem{Giacometti1995}
Giacometti A 1995
Exact closed form of the return probability on the Bethe lattice
{\it J. Phys. A} {\bf 28} L13

\bibitem{Hughes1982}
Hughes B D and Sahimi M 1982
Random walks on the Bethe lattice
{J. Stat. Phys.} {\bf 29} 781

\bibitem{Cassi1989}
Cassi D 1989
Random walks on Bethe lattices
{\it Europhys. Lett.} {\bf 9} 627


\bibitem{Tishby2021b}
Tishby I, Biham O and Katzav E 2021
Analytical results for the distribution of  
first return times of random walks on random regular graphs
{\it J. Phys. A} {\bf 54} 325001


\bibitem{Molloy1995}
Molloy M and Reed A 1995 
A critical point for random graphs with a given degree sequence
{\it Rand. Struct. Alg.} {\bf 6}  161-180 


\bibitem{Molloy1998}
Molloy M and Reed A 1998 
The Size of the Giant Component of a Random Graph with a Given Degree Sequence
{\it Combinatorics, Probability and Computing} {\bf 7}  295-305 


\bibitem{Newman2001}
Newman M E J, Strogatz S H and Watts D J 2001
Random graphs with arbitrary degree distributions and their applications,
{\it Phys. Rev. E} {\bf 64}, 026118 (2001).



\bibitem{Bollobas2001}
Bollobas B 2001
{\it Random Graphs, Second Edition}
(London: Academic Press)



\bibitem{Bonneau2017}
Bonneau H, Hassid A, Biham O, K\"uhn R and Katzav E 2017 
Distribution of shortest cycle lengths in random networks 
{\it Phys. Rev. E} {\bf 96} 062307 


\bibitem{Nitzan2016}
Nitzan M, Katzav E, K\"uhn R and Biham O 2016,
Distance distribution in configuration-model networks
{\it Phys. Rev. E}{\bf 93}  062309 
 
 
\bibitem{Hofstad2005}
van der Hofstad R, Hooghiemstra G and  Van Mieghem P 2005
Distances in random graphs with finite variance degrees
{\it Random Structures \& Algorithms}
{\bf 27} 76  
 
 
\bibitem{Gompertz1825}
Gompertz B 1825 
On the nature of the function expressive of the law of human mortality, 
and on a new mode of determining the value of life contingencies
Philosophical Trans. R. Soc. London A 115 513


\bibitem{Shklovskii2005}
Shklovskii B I 2005 
A simple derivation of the Gompertz law for human mortality
{\it Theory in Biosciences} {\bf 123} 431



\bibitem{Tishby2022}
Tishby I, Biham O, K\"uhn R and Katzav E 2022
The mean and variance of the distribution of
shortest path lengths of random regular graphs
{\it J. Phys. A} {\bf 55} 265005 





\bibitem{Olver2010}
Olver F W J, Lozier D M, Boisvert R R and Clark C W 2010
{\it NIST Handbook of Mathematical Functions}
(Cambridge: Cambridge University Press)





\bibitem{Tishby2022b}
Tishby I, Biham O and Katzav E 2022
Analytical results for the distribution of  
cover times of random walks on random regular graphs
{\it J. Phys. A} {\bf 55} 015003



\bibitem{Bollobas1982}
Bollobas B and Fernandez de la Vega W 1982
The diameter of random regular graphs 
{\it Combinatorica} {\bf 2} 125



\bibitem{Pitman1993}
Pitman J 1993
{\it Probability} 
(New York: Springer-Verlag)



\bibitem{Martin2010}
Martin O and Šulc P 2010
Return probabilities and hitting times of random walks on sparse Erd{\H o}s-R\'enyi graphs
{\it Phys. Rev. E} {\bf 81} 031111



\bibitem{Pitman2018}
Pitman J and Tang W 2018
Tree formulas, mean first passage times and Kemeny’s constant of a Markov chain
{\it Bernoulli} {\bf 24}, 1942

\bibitem{Forster2022a}
F\"orster Y-P, Gamberi L, Tzanis E, Vivo P and Annibale A 2022
Exact and approximate mean first passage times on trees and other necklace
structures: a local equilibrium approach
{\it J. Phys. A} {\bf 55}, 115001




\bibitem{Forster2022b}
F\"orster Y-P, Annibale A, Gamberi L, Tzanis E and Vivo P 2022 
Information retrieval and
structural complexity of legal trees
{\it J. Phys. Complex.} {\bf 3}, 035008






\bibitem{Tishby2017}
Tishby I, Biham O and Katzav E 2017
The distribution of first hitting times of
random walks on Erdős–Rényi networks
{\it J. Phys. A}, {\bf 50} 115001





\bibitem{Tishby2021a}
Tishby I, Biham O and Katzav E 2021
Analytical results for the distribution of first hitting times
of random walks on random regular graphs
{\it J. Phys. A}  {\bf 54} 145002


\bibitem{Cooper2005}
Cooper C and Frieze A M 2005
The cover time of random regular graphs
{\it SIAM J. Discrete Math.} {\bf 18} 728 










\bibitem{Gumbel1935}
Gumbel E J 1935 
Les valeurs extr\^emes des distributions statistiques
{\it Annales de l'Institut Henri Poincar\'e} {\bf 5} 115 



\bibitem{Frechet1927}
Fr\'echet M 1927
Sur la loi de probabilit\'e de l'\'ecart maximum
{\it Annales de la Soci\'et\'e Polonaise de Math\'ematique}
{\bf 6} 93 
 
\bibitem{Fisher1928}
Fisher R A, Tippett L H C 1928
Limiting forms of the frequency distribution of the largest and smallest member of a sample
{\it Proc. Camb. Phil. Soc.} {\bf 24} 180 


\bibitem{Mises1936}
von Mises R 1936
La distribution de la plus grande de $n$ valeurs
{\it Rev. Math. Union Interbalcanique} {\bf 1} 141 


\bibitem{Gnedenko1943}
Gnedenko B V 1943
Sur la distribution limite du terme maximum d'une serie aleatoire
{\it Annals of Mathematics} {\bf 44} 423 




\bibitem{Audibert2010}
Audibert P 2010 
{\it Mathematics for Informatics and Computer Science}
(Hoboken: Wiley-ISTE).




\bibitem{Erdelyi1953}
Erdelyi A (ed.) 1953
{\it Higher Transcendental Functions, Vol. I}
(New York: McGraw-Hill)


\end{thebibliography}
\end{document}